\documentclass[useAMS,graphics,usenatbib]{mn2e}
\usepackage{amssymb,color}
\usepackage{epsfig}

\newcommand{\traphic}{\textsc{TRAPHIC}}

\newcommand{\erg}{~\mbox{erg}}

\newcommand{\eV}{~\mbox{eV}}

\newcommand{\cMpch}{~h^{-1}~\mbox{comoving Mpc}}
\newcommand{\Mpch}{~h^{-1}~\mbox{Mpc}}
\newcommand{\kpch}{~h^{-1}~\mbox{kpc}}
\newcommand{\ckpch}{~h^{-1}~\mbox{comoving kpc}}

\newcommand{\cmsqi}{~\mbox{cm}^{-2}}
\newcommand{\cmci}{~\mbox{cm}^{-3}}

\newcommand{\K}{~\mbox{K}}

\newcommand{\invyr}{~\mbox{yr}^{-1}}

\newcommand{\Msunh}{~h^{-1}~\mbox{M}_{\odot}}
\newcommand{\Msun}{~\mbox{M}_{\odot}}
\newcommand{\Zsun}{~Z_{\odot}}

\newcommand{\cmsi}{~\mbox{cm}^{-2}}

\title[Aurora]{The Aurora radiation-hydrodynamical simulations of reionization: calibration and first results.}

\author[Pawlik et al.]
{Andreas H. Pawlik$^{1}$\thanks{E-mail:pawlik@mpa-garching.mpg.de},
  Alireza Rahmati$^2$, Joop Schaye$^3$, Myoungwon Jeon$^4$, 
\newauthor
Claudio Dalla Vecchia$^{5,6}$\\
   $^1$Max-Planck Institute for Astrophysics, Karl-Schwarzschild-Strasse 1, 85748 Garching, Germany\\
   $^2$Institute for Computational Science, University of Z\"urich, Winterthurerstrasse 190, CH-8057 Z\"urich, Switzerland\\
   $^3$Leiden Observatory, Leiden University, P.O. Box 9513, 2300 RA
   Leiden,The Netherlands\\
  $^4$Department of Astronomy and Steward Observatory, University of
  Arizona, Tucson, AZ 85721-0065, USA\\
   $^5$Instituto de Astrof\'isica de Canarias, C/ V\'ia L\'actea s/n, 38205 La Laguna, Tenerife, Spain \\
   $^6$Departamento de Astrof\'isica, Universidad de La Laguna, Av. del Astrof\'isico Fracisco S\'anchez s/n, 38206 La Laguna, Tenerife, Spain}

\begin{document}

\date{}

\pagerange{\pageref{firstpage}--\pageref{lastpage}} \pubyear{xxxx}

\maketitle

\label{firstpage}

\begin{abstract}
We introduce a new suite of radiation-hydrodynamical simulations of
galaxy formation and reionization called Aurora. The Aurora
simulations make use of a spatially adaptive radiative transfer
technique that lets us accurately capture the small-scale structure in
the gas at the resolution of the hydrodynamics,
in cosmological volumes.  In addition to ionizing
radiation, Aurora includes galactic winds driven by star formation and the
enrichment of the universe with metals synthesized in the stars.  Our
reference simulation uses $2\times 512^3$ dark matter and gas
particles in a box of size $25 \cMpch$ with a force softening scale of
at most $0.28 \kpch$. It is accompanied by simulations in larger and
smaller boxes and at higher and lower resolution, employing up to $2\times 1024^3$ particles,  to investigate
numerical convergence. All simulations are calibrated to yield
simulated star formation rate (SFR) functions in close agreement with
observational constraints at redshift $z = 7$ and to achieve
reionization at $z \approx 8.3$, which is consistent with the observed
optical depth to reionization. We focus on the
design and calibration of the simulations and present some first
results. The median stellar metallicities of low-mass galaxies at $z = 6$ are
consistent with the metallicities of dwarf galaxies in the Local
Group, which are believed to have formed most of their stars at high
redshifts. After reionization, the mean photoionization rate decreases systematically with increasing resolution. This coincides with a systematic increase in the abundance of neutral hydrogen absorbers in the IGM.

\end{abstract}


\section{Introduction}
Reionization is the transformation of the initially cold and neutral gas 
which resulted from recombination after the Big Bang into the hot and ionized plasma that we
observe between galaxies today. Measurements of the electron scattering
optical depth from observations of the cosmic microwave background
(CMB; e.g., \citealp{Planck2015}) and the detection of extended
Gunn-Peterson absorption troughs in the spectra of high-redshift
quasars (e.g., \citealp{Becker2001}; \citealp{Fanreview2006}) have
firmly established that reionization took place in the first billion
years. This is a key epoch in the history of the Universe that has also seen
the birth of the first stars, the initial enrichment of the cosmic gas with
metals synthesized in the stars, and the formation of galaxies,
including the progenitors of more massive galaxies like our
own galaxy, the Milky Way (for reviews see, e.g., \citealp{Barkana2001}; 
\citealp{Ricotti2010}; \citealp{Bromm2011}; \citealp{Zaroubi2013};
\citealp{Lidz2015}, \citealp{McQuinn2015}). 
\par 
Understanding the physics of reionization is a primary goal of modern
astrophysics. One of the main open questions that drives research in
the field is whether the ionizing radiation from stars was
sufficient to accomplish reionization, or if there were other, perhaps
yet to be discovered, sources at play (e.g., \citealp{Volenteri2009};
\citealp{Loeb2009}; \citealp{Madau2015}).  Estimates from
observations of the early Universe with the Hubble Space Telescope are
consistent with a picture in which reionization is driven mainly by
star-forming galaxies (e.g., \citealp{Pawlikclump2009}; \citealp{Fontanot2012};
\citealp{Finkelstein2014}; \citealp{Sharma2015}). However, such
estimates still require assumptions about, e.g., the
escape fraction of ionizing radiation (e.g., \citealp{Wise2009};
\citealp{Yajima2011}; \citealp{Paardekooper2015}), absorption of
ionizing photons at small
scales (e.g., \citealp{Finlatorclump2012}; \citealp{Kaurov2014};
\citealp{Pawlik2015}, hereafter PSD15) and the abundance of faint galaxies below the
current detection threshold, all of which are difficult to test
observationally (for a recent review see \citealp{BouwensReview2015}).
\par 
Upcoming observations of galaxies with, e.g., ALMA (e.g.,
\citealp{Ota2014}), MUSE (e.g., \citealp{Wisotzki2015}) and JWST (e.g.,
\citealp{Finkelstein2015}), and of the
intergalactic medium (IGM) with, e.g., LOFAR (e.g.,
\citealp{Zaroubi2012}), MWA (e.g., \citealp{Lidz2008}) and SKA (e.g.,
\citealp{Mellema2013}), 
promise to shed further light on these issues. Theoretical models of
the early Universe will be critical 
to help interpret the data that these telescopes will collect. Cosmological
radiative transfer (RT) simulations of reionization have emerged as some of the
most powerful approaches to building such models. These simulations
track the gravitational growth of density fluctuations in the
expanding Universe and the hydrodynamical evolution of the
cosmic gas, include recipes for star formation and the associated feedback, and also
follow the propagation of ionizing radiation (e.g.,
\citealp{Ciardi2003}; \citealp{Iliev2006}; \citealp{McQuinn2007};
\citealp{Petkova2011}; \citealp{Hasegawa2013};
\citealp{Gnedin2014}; PSD15; \citealp{Aubert2015}). The main drawback of these
simulations is the extreme computational challenge that they pose, in particular the accurate transport of ionizing photons from multiple
sources across a large range of scales in cosmological volumes  (for
reviews see, e.g., \citealp{Trac2011};
\citealp{Finlator2012}; \citealp{Dale2015}).
\par  
In this work we introduce Aurora\footnote{Aurora is the Latin word for dawn,
  and the goddess of dawn in Roman mythology and Latin poetry.}, a new suite of cosmological radiation-hydrodynamical
Smoothed Particle Hydrodynamics (SPH) simulations of reionization. The Aurora simulations track the formation of
stars and galaxies at high resolution across cosmological scales, and also
follow the hydrodynamically coupled transport of ionizing radiation from stars, the
explosion of stars as supernovae (SNe) and the enrichment of the universe with
metals.
\par 
Aurora addresses the computational challenges imposed by the RT by
using the spatially adaptive radiation transport technique \traphic\
(\citealp{Pawlik2008}) that solves multi-scale problems efficiently and has a computational
cost that is independent of the number of radiation sources. Ionizing
radiation is transported at the same high spatially adaptive
resolution at which the gas is evolved. The ionizing radiation is
coupled hydrodynamically to the evolution of the gas, which lets us
accurately account for feedback from
photoheating. 
\par
For comparison, in most
previous reionization simulations the radiation transport was carried
out on a uniform grid superimposed on the
cosmological simulation (e.g., \citealp{Iliev2006};
\citealp{Finlator2011}; \citealp{Ciardi2012}; \citealp{Iliev2014}). Because computational
resources are finite, such a grid
typically implies a substantial reduction in dynamic range in
comparison with the underlying spatially adaptive gas simulation and
impedes the coupling of the radiation and the dynamics of the gas. In
fact, the gas dynamics has
often been ignored altogether, assuming the gas traces the dark matter, thus preventing an
accurate treatment of the feedback from photoionization heating. Only 
recently have cosmological simulations started to approach the scales 
and the high resolution needed to capture the relevant
physics (e.g., \citealp{Gnedin2014}; \citealp{Norman2015};
PSD15; \citealp{Bauer2015}; \citealp{Ocvirk2015}).
\par
Simulating the internal structure of galaxies in cosmological volumes
from first principles, is currently not computationally feasible. This holds even for substantially
less expensive simulations in which the accurate transfer of ionizing radiation
is ignored.  For this reason, cosmological simulations like ours must make
use of subresolution models to incorporate processes that are not resolved
but which are known to impact the dynamics at resolved scales. Because the
coupling between unresolved and resolved physics is often not sufficiently
understood, this introduces free parameters.  Two parameters that are
especially relevant to simulations of galaxy formation during reionization are the
efficiency of stellar feedback due to the injection of energy by
stellar winds and by SN explosions, and the subresolution escape fraction of ionizing radiation.
\begin{figure*}
\centerline{\hbox{\includegraphics[width=0.5\textwidth]
             {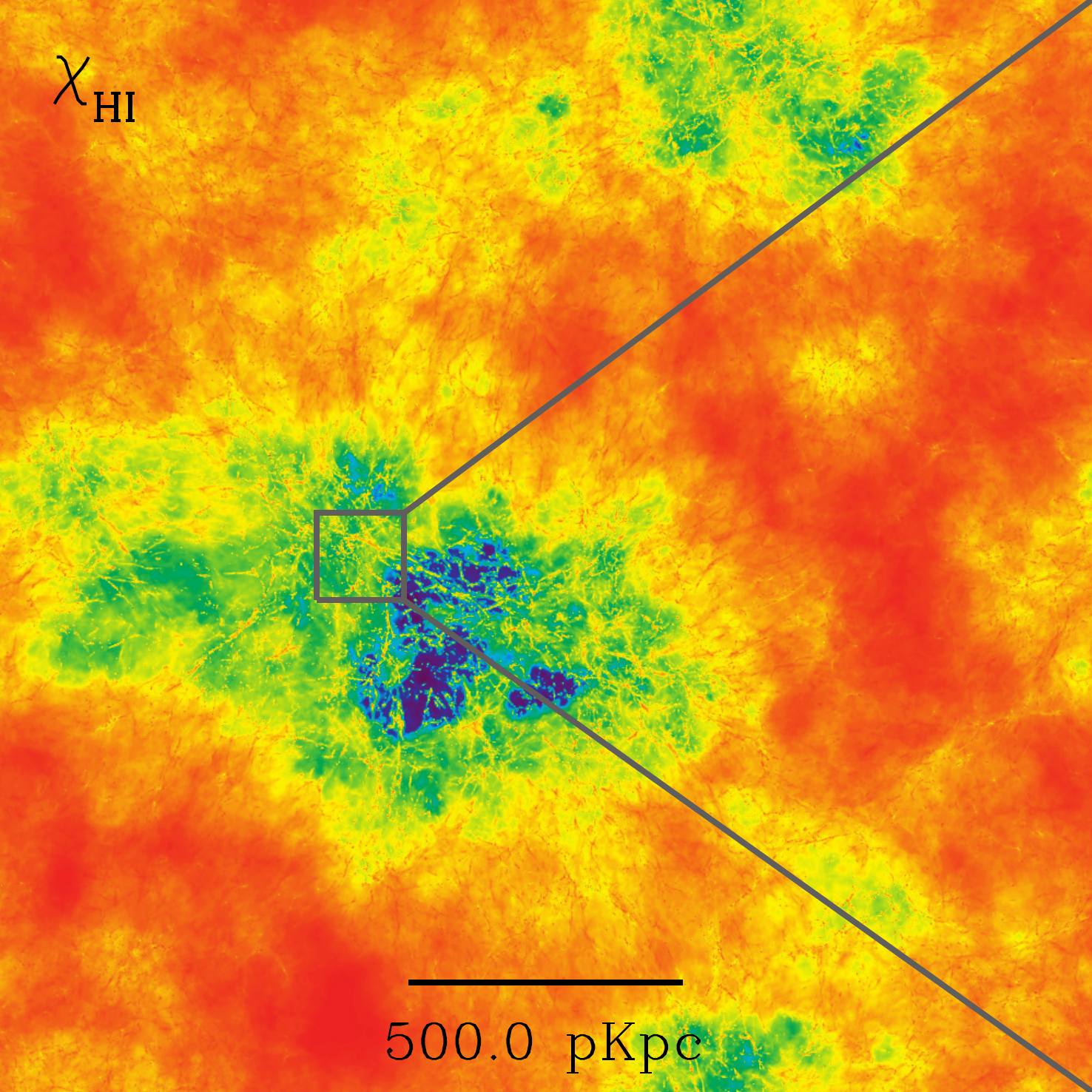}}
             \hbox{\includegraphics[width=0.5\textwidth]
             {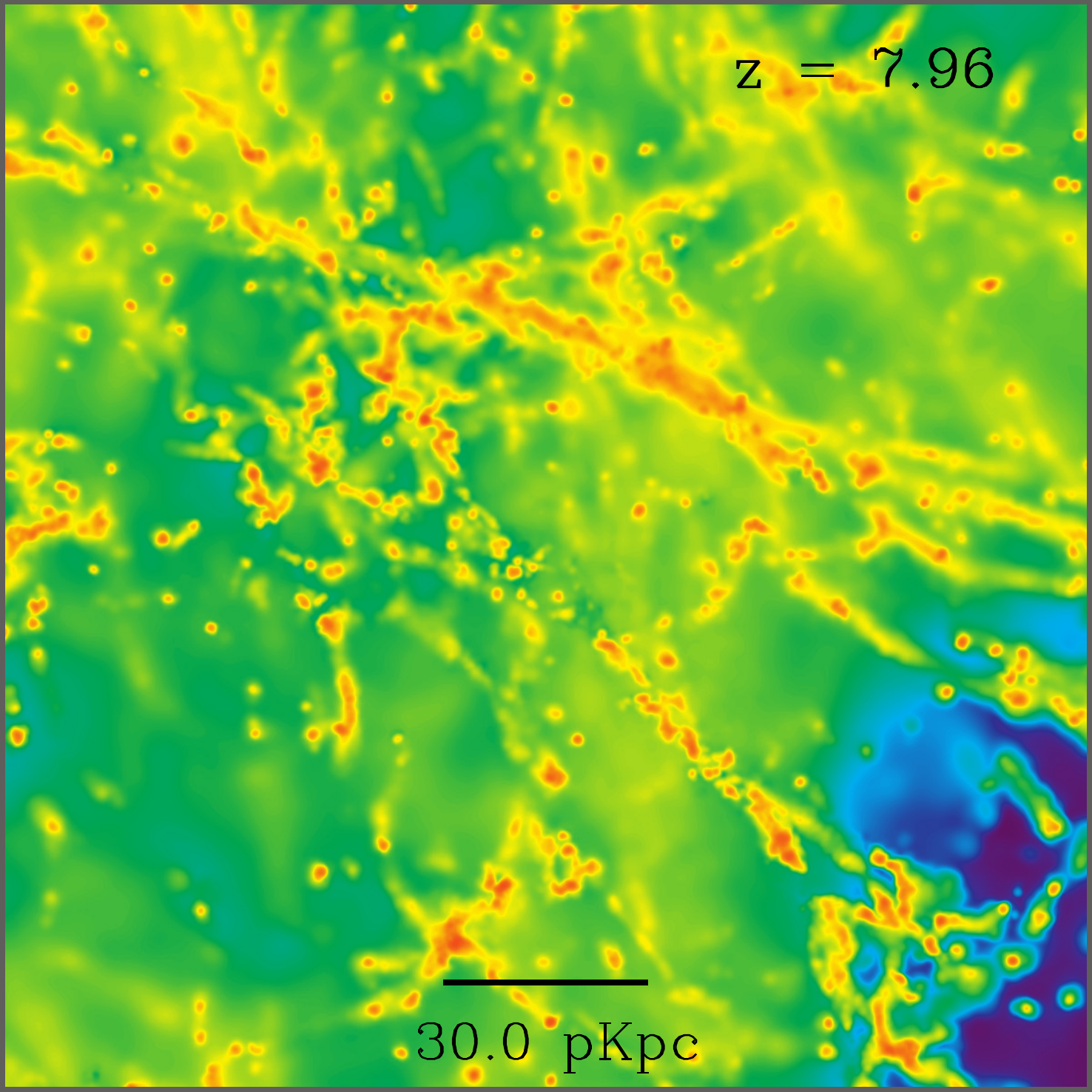}}}
\centerline{\hbox{\includegraphics[width=0.8\textwidth]
             {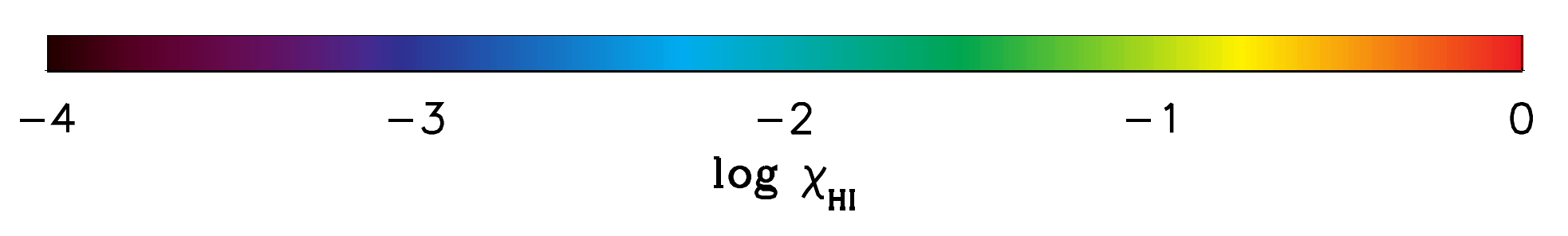}}}

\caption{The mass-weighted neutral hydrogen fraction at $z \approx 8$,
  corresponding to a mean volume-weighted neutral hydrogen fraction of
  about 0.4,  in our highest resolution simulation L012N0512 that used $2 \times 512^3$
  dark matter and gas particles in a box of linear extent $12.5
  \cMpch$. The zoom into the selected region (right panel) reveals a
  network of self-shielded and shadowed neutral gaseous objects inside
  the highly ionized regions and demonstrates a large dynamic range in
  the neutral density captured by our spatially adaptive radiative
  transfer technique. \label{fig:zoom}}

\end{figure*}

\par
In Aurora, we calibrate the efficiency of stellar feedback and the
subresolution escape fraction to match current observational constraints on 
the SFR function at $z \ge 6$ and the redshift of reionization. The calibration is
carried out for a set of simulations that cover a range of box sizes
and resolutions. This lets us investigate the numerical convergence of
our simulations at fixed reionization redshift, which removes a major
variable that may complicate the interpretation of convergence
tests. Quantities that have not been calibrated, such as the
slope of the SFR function at SFRs that are lower than currently observed, the time-dependent distribution
of neutral gas, and the strength of the photoionizing background that
fills space after reionization, are direct predictions coming out of
Aurora. Calibration is key to embedding these predictions in
observationally supported models of reionization 
\par
Our approach follows that of \cite{Schaye2015}, who introduced the
notions of strong and weak convergence. Simulations converge strongly
if the simulation outcome does not change significantly when
the resolution is increased while holding all other parameters fixed. Strong convergence is
computationally expensive to achieve and might still say little about the 
quality of the simulations. This is because subresolution models
are typically only applicable within a
limited range of resolutions. Weak convergence
acknowledges these difficulties by demanding convergence only after
an appropriate adaption of the subresolution model. In other words, 
model parameters that are critical for the outcome but whose values cannot be 
predicted from first principles should be calibrated, using observations if possible.
\par
\cite{Gnedin2016} also investigated the weak and strong convergence of the SFR 
in simulations of reionization. Our approach
differs from that work because we calibrate subresolution model
parameters to match the observed SFR function rather than the
theoretically estimated SFR function, which \cite{Gnedin2016} 
determines by extrapolating the results from a series of simulations
of increasing resolution. In addition, unlike \cite{Gnedin2016}, we calibrate also
the subresolution escape fraction, which is needed for achieving weak
convergence in the reionization redshift, the other major observable
of the epoch of reionization that currently cannot be predicted from first
principles.
\par
This paper focuses on the design of the Aurora simulations and
discusses results from the initial analysis of a subset of the
simulations carried out in the Aurora programme. A more comprehensive
and detailed analysis of the simulations will be presented in future
work. There we will also discuss additional simulations, for which we
have varied the numerical parameters away from their calibrated values
to help gain further insight in the simulation results (similar to
what we did in PSD15).
\par
The Aurora simulations start from the same
initial conditions as the spatially adaptive radiation-hydrodynamical
reionization simulations presented in PSD15. They complement the
simulations in PSD15 by means of
calibration, which we use to remove the dependence on resolution of the
comparison between simulated and observed SFR functions and of the redshift
of reionization, two key observables of the early Universe. 
They improve on them by the increased physical realism, additionally
including, e.g., absorption of ionizing photons by helium and chemical
enrichment, and by the use of a larger range of box sizes and more particles.
\par
The structure of this paper is as follows. In Section 2 we describe our
numerical techniques and the set of simulations that we investigate here. In
Section 3 we describe the calibration of the efficiency of the stellar
feedback and of the escape fraction of ionizing photons. In Section 4 we
describe a set of initial results, including the star formation
and reionization history, the build-up of the ionizing
background and metal enrichment. We also discuss the
properties of the first galaxies. In Section 5 we conclude with a
brief summary. Lengths are expressed in physical units, unless noted
otherwise.

\section{Simulations}

\begin{table*}
\begin{center}
\caption{Simulation parameters:
  simulation name;
  comoving size of the simulation box, $L_{\rm box}$;
  number of DM particles, $N_{\rm DM}$;
  mass of dark matter particles, $m_{\rm DM}$;
  initial mass of gas particles, $m_{\rm gas}$;
  gravitational softening, $\epsilon_{\rm soft}$;
  stellar feedback efficiency, $f_{\rm SN}$;
  subresolution ISM escape fraction of ionizing photons, $f^{\rm subres}_{\rm esc}$;
  comment. The number of SPH particles initially equals $N_{\rm DM}$
  (and it decreases
  during the simulation due to star formation). The mass of SPH particles is
  initially lower than the mass of DM particles by a factor $\Omega_{\rm DM} /
  \Omega_{\rm b} \approx 4.9$. The mass of SPH particles may increase above this
  initial value as stellar winds and SNe transfer stellar material to the gas.
  \label{tbl:params}}

\begin{tabular}{lrrrrrrrlll}

\hline
\hline  
simulation & $L_{\rm box}$ & $N_{\rm DM}$ & $m_{\rm DM}$  &$m_{\rm gas}$
&$\epsilon_{\rm soft}$ & $f_{\rm SN}$ & $f_{\rm esc}^{\rm subres}$ & comment &\\
           & $[\cMpch]$    &              & $[\Msun]$  & $[\Msun]$& $[\ckpch]$    &\\           
\hline

{L025N0512}        & $25.0$ & $512^3$ & $1.00 \times 10^7$ &$2.04 \times 10^6$ & 1.95  &0.8&0.6&  reference &\\

{L012N0256}        & $12.5$ & $256^3$ & $1.00 \times 10^7$ &$2.04 \times 10^6$ & 1.95 &0.8&0.6&   small box &  \\

{L012N0512}        & $12.5$ & $512^3$ & $1.25 \times 10^6$ &$2.55 \times 10^5$ &0.98 &0.6& 0.5 &  high resolution   &  \\

{L025N0256}        & $25.0$ & $256^3$ & $8.20 \times 10^7$ &$1.63 \times 10^7$ & 3.91 &1.0&1.3&    low resolution &  \\
{L050N0512}        & $50.0$ & $512^3$ & $8.20 \times 10^7$ &$1.63 \times 10^7$ & 3.91 &1.0&1.3&  large box &  \\

{L100N1024}        & $100.0$ & $1024^3$ & $8.20 \times 10^7$ &$1.63 \times 10^7$
& 3.91 &1.0&1.3& $z_{\rm current} = 8.4$ &  \\

\hline
\label{tab:sims}
\end{tabular}

\end{center}
\end{table*}

We use a modified version of the N-body/TreePM Smoothed Particle Hydrodynamics
(SPH) code GADGET (last described in \citealp{Springel2005}) to perform a
suite of cosmological radiation-hydrodynamical simulations of galaxy formation
down to redshift $z = 6$ (see Table~\ref{tab:sims} and Figure~\ref{fig:zoom}). The simulations presented
 here start at $z = 127$ from the same initial conditions as the simulations published in
PSD15. However, the two suites of
simulations differ substantially in a number of points.
\par
A major difference is that in the current work we calibrate the stellar feedback 
strength and the unresolved escape fraction of ionizing radiation separately for each of the simulations. Consequently, 
all simulations of the current suite yield nearly identical SFR
functions and reionization
histories, independent of box size and numerical resolution and consistent
with observations. This is complementary to our previous work, where the
parameters were identical for all simulations and the match between simulated
and observed SFR functions and the redshift of reionization both
depended on resolution.
\par
Other differences include that in this work we follow the chemistry
and cooling of helium in addition to that of hydrogen, including ionization 
of helium by stellar radiation. New is also that we track the
enrichment with metals of the IGM. Finally, the current work makes use
of an improved implementation of stellar feedback and also puts stricter
limits on the size of the time steps over which the dynamics of the
SPH particles are integrated, which improves the accuracy at which the
hydrodynamics equations are solved. 
\par
In the following summary of our methodology we will focus
on the differences with respect to PSD15, to which we refer the
reader for details. We adopt the $\Lambda$CDM cosmological model with
parameters 
$\Omega_{\rm m} = 0.265$, $\Omega_{\rm b} = 0.0448$ and 
$\Omega_{\Lambda} = 0.735$, $n_{\rm s} = 0.963$, $\sigma_8 = 0.801$,
and $h = 0.71$ (\citealp{Komatsu2011}), consistent with the most recent constraints from observations
of the CMB by the {\it Planck} satellite (\citealp{Planck2015}).

\subsection{Hydrodynamics}
As in PSD15, we adopt the 
entropy-conserving formulation of SPH (\citealp{Springel2005}) 
and average over 48 SPH neighbor particles. 
New is that we use the time step limiter described in \cite{Durier2012} to keep
the ratio of time steps of neighboring SPH particles $\le 4$. This 
ensures an accurate energy-conserving integration of the equations of 
hydrodynamics (see also \citealp{Saitoh2009}; \citealp{Springel2010}). 

\subsection{Chemistry and cooling} 
We follow the 
non-equilibrium chemistry and radiative cooling of the gas, assuming 
that the gas is of primordial composition with a hydrogen mass
fraction $X = 0.75$ and a helium mass fraction $Y=1-X$.  This is a substantial
improvement on PSD15, in which we neglected helium to speed up the
simulations. We achieve this improvement by replacing the explicit chemistry
solver described in PSD15 with the implicit solver
described in \cite{Pawlik2013}, which is faster. As in PSD15, we
impose a temperature floor above densities $n_{\rm H}
\ge 0.1 \cmci$ to prevent spurious fragmentation in dense underresolved gas, following \cite{Schaye2008}.

\subsection{Star formation} 
Our implementation of star formation is identical to that used in
PSD15. This means that we use the pressure-dependent stochastic recipe of
\cite{Schaye2008} to sample from a \cite{Chabrier2003} initial mass function (IMF) in the
range $0.1-100 \Msun$ and turn gas particles with densities above $n_{\rm
  H}\ge 0.1 \cmci$ into star particles, at a rate that reproduces the
observed star formation law in simulations of isolated disk galaxies
(see \citealp{Schaye2008}  for a detailed discussion).

\subsection{Chemical enrichment}
Another major improvement on PSD15 is that we account 
for the age-dependent release of hydrogen, helium, and metals by Type II core-collapse
and Type Ia SNe, stellar winds from massive stars, and AGB stars, following
\cite{Schaye2010} and \cite{Wiersma2009}. We track both the total ejected metal mass, and,
separately, the ejected mass of 11 individual elements, including H,
He, C, N, O, Ne, Mg, Si, and Fe.  Unlike \cite{Schaye2010}, we use particle metallicities,
not SPH smoothed metallicities, and we neglect metal-line cooling (see, e.g., \cite{Schaye2015} 
for a discussion of the impact on metal line cooling on the calibration of SFR functions). Chemical
enrichment thus impacts our simulations solely by setting the metallicity-dependent luminosities of the star particles (\citealp{Schaerer2003}) and
the duration of the various phases of stellar evolution, determining 
the mass loss from stellar winds and the timing of SN explosions (\citealp{Wiersma2009}).

\begin{figure*}
  \begin{center}

    \includegraphics[width=0.49\textwidth,clip=true, trim=0 0 0 0,
      keepaspectratio=true]{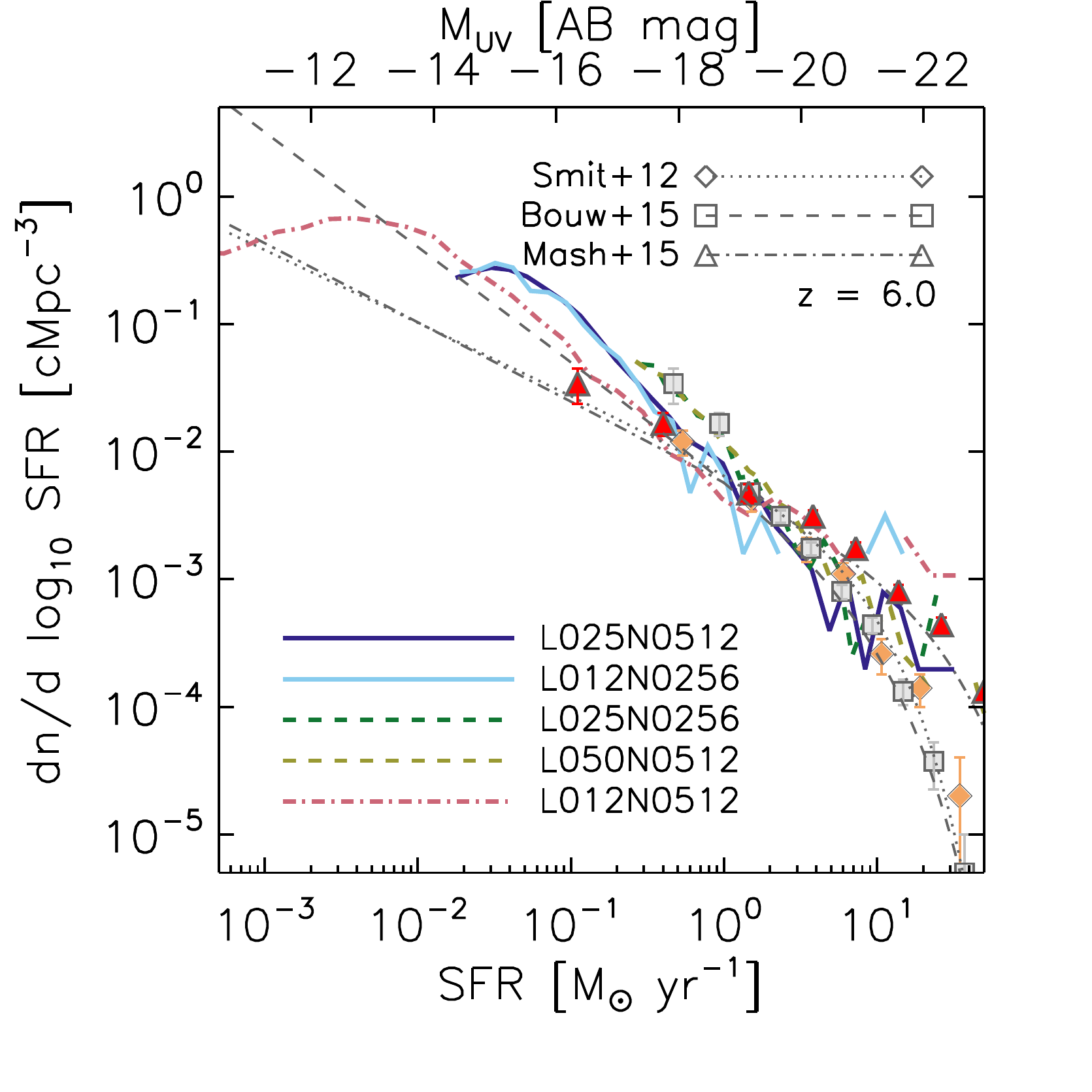}
    \includegraphics[width=0.49\textwidth,clip=true, trim=0 0 0 0,
      keepaspectratio=true]{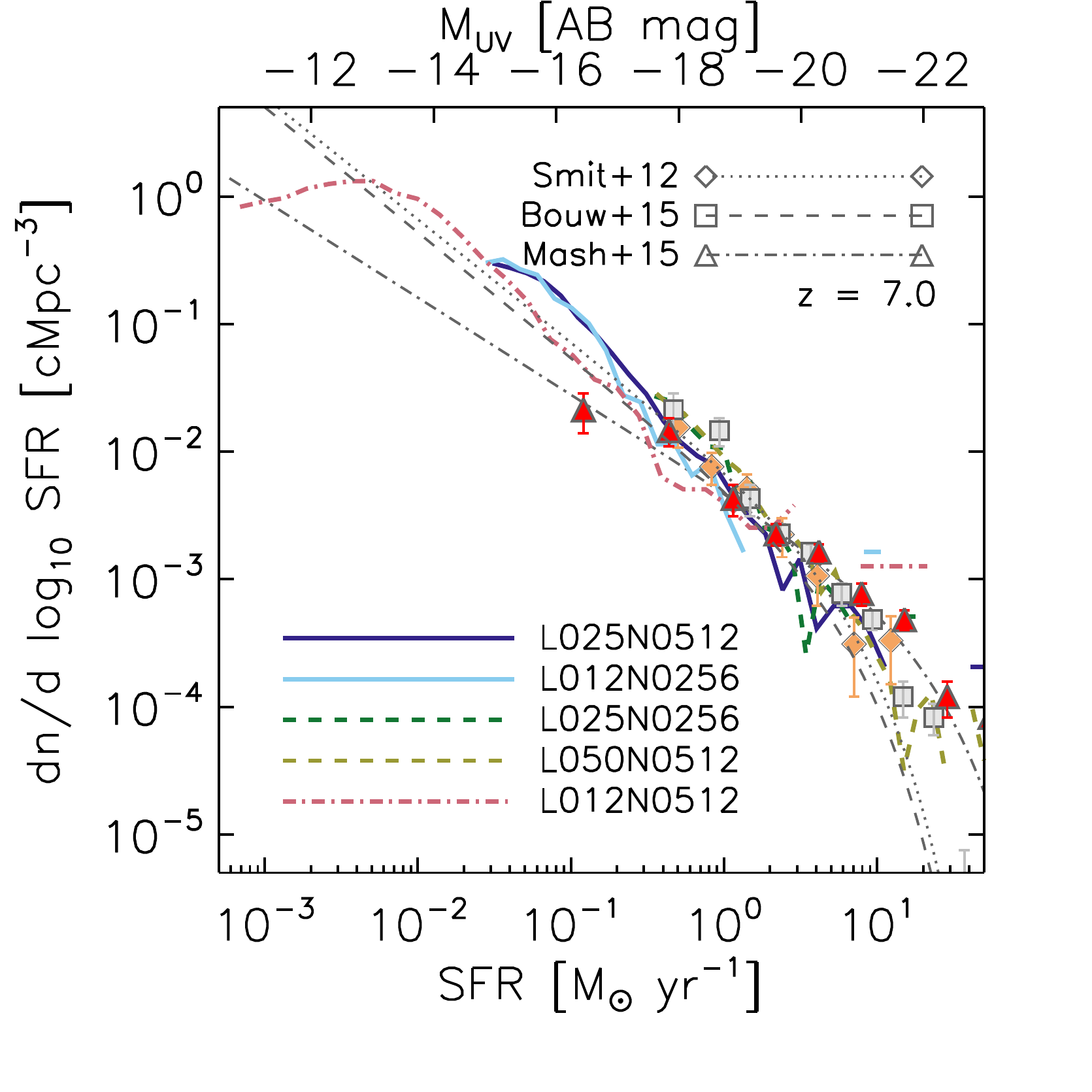}
    \\
    \includegraphics[width=0.49\textwidth,clip=true, trim=0 0 0 0,
      keepaspectratio=true]{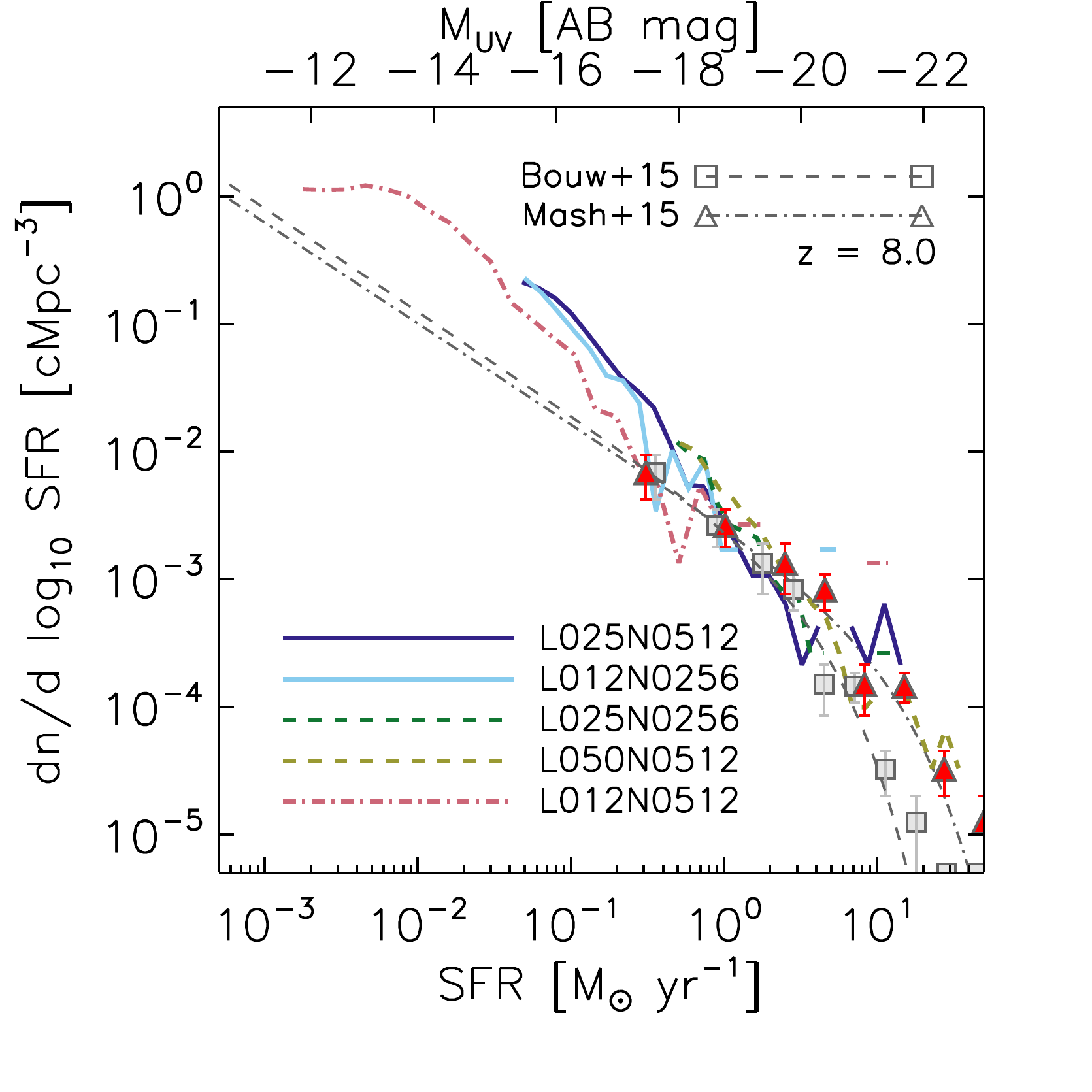}
    \includegraphics[width=0.49\textwidth,clip=true, trim=0 0 0 0,
      keepaspectratio=true]{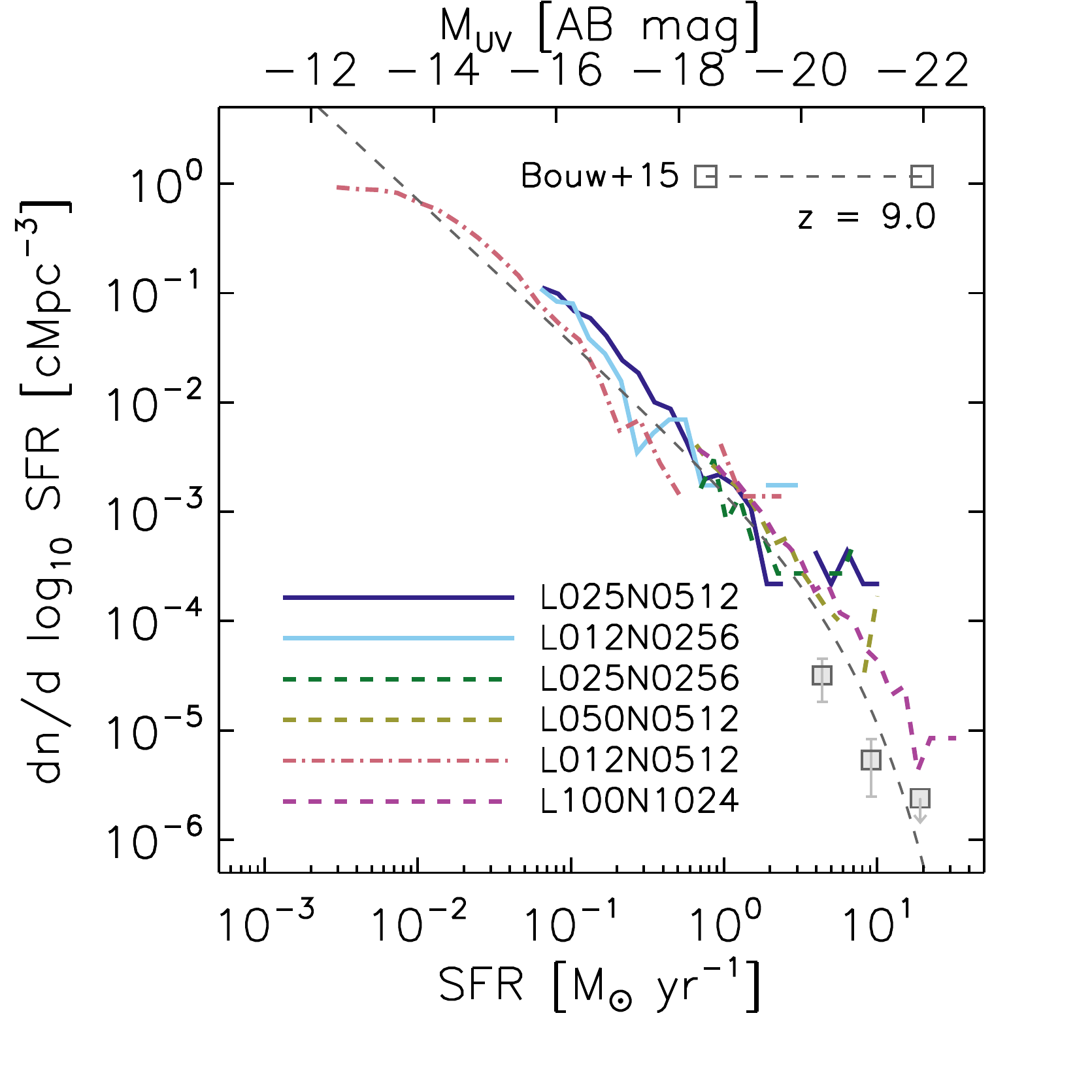}

  \end{center}
  \caption{The SFR functions at $z = 6, 7, 8$, and 9 for
    the full set of simulations listed in Table~\ref{tab:sims}, as indicated in the legends. The top axis shows
    the corresponding UV luminosity, computed as in
     \protect\cite{Kennicutt1998} and converted to the Chabrier IMF. Symbols show
    recent observational constraints from \protect\citet[diamonds,
    corrected for dust]{Smit2012}, \protect\citet[squares, no dust correction]{Bouwens2015},
    \protect\citet[triangles, corrected for dust]{Mashian2015}, and
    the grey
    curves show the corresponding Schechter fits. Simulation L100N1024 has been stopped at $z = 8.4$ and
    is therefore only included in the comparison at $z = 9$. The
    simulated SFR function is only shown at SFRs larger than the median SFR of halos
    with mass larger than 100 DM particles. Where
    necessary, we have converted the observed
    SFRs to the Chabrier IMF. We have
    calibrated the simulations at $z = 7$ by adopting a stellar
    feedback efficiency parameter $f_{\rm SN}$
    that yields close agreement between the simulated SFR
    functions and those observationally inferred by \protect\cite{Smit2012}. The
    simulations yield SFR functions in close agreement with the
    observational constraints over the range of observed SFRs at $z\gtrsim 6$.}
  \label{fig1}
\end{figure*}

\begin{figure*}
  \begin{center}

    \includegraphics[width=0.49\textwidth,clip=true, trim=0 0 0 0,
      keepaspectratio=true]{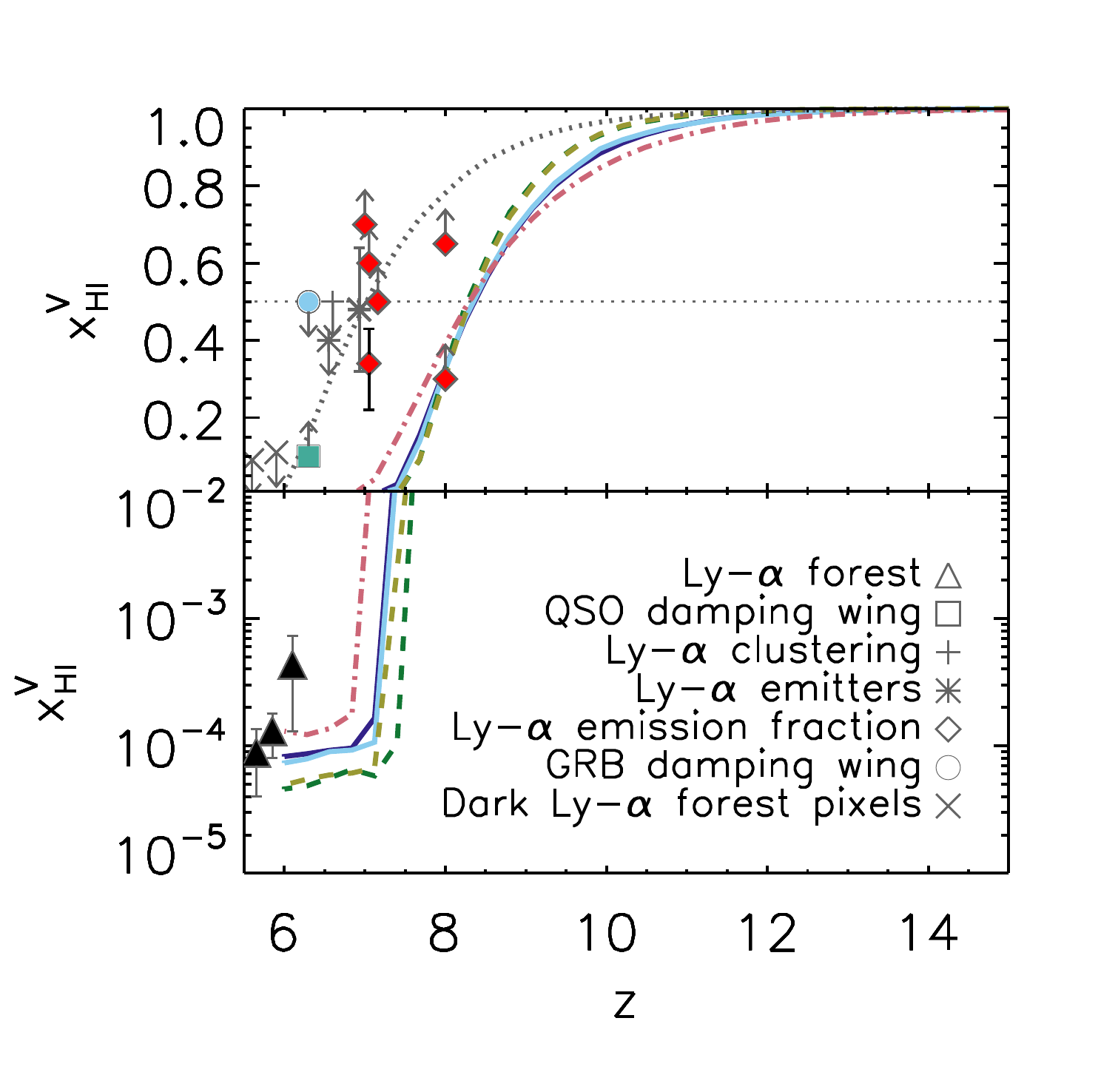}
    \includegraphics[width=0.49\textwidth,clip=true, trim=0 0 0 0,
      keepaspectratio=true]{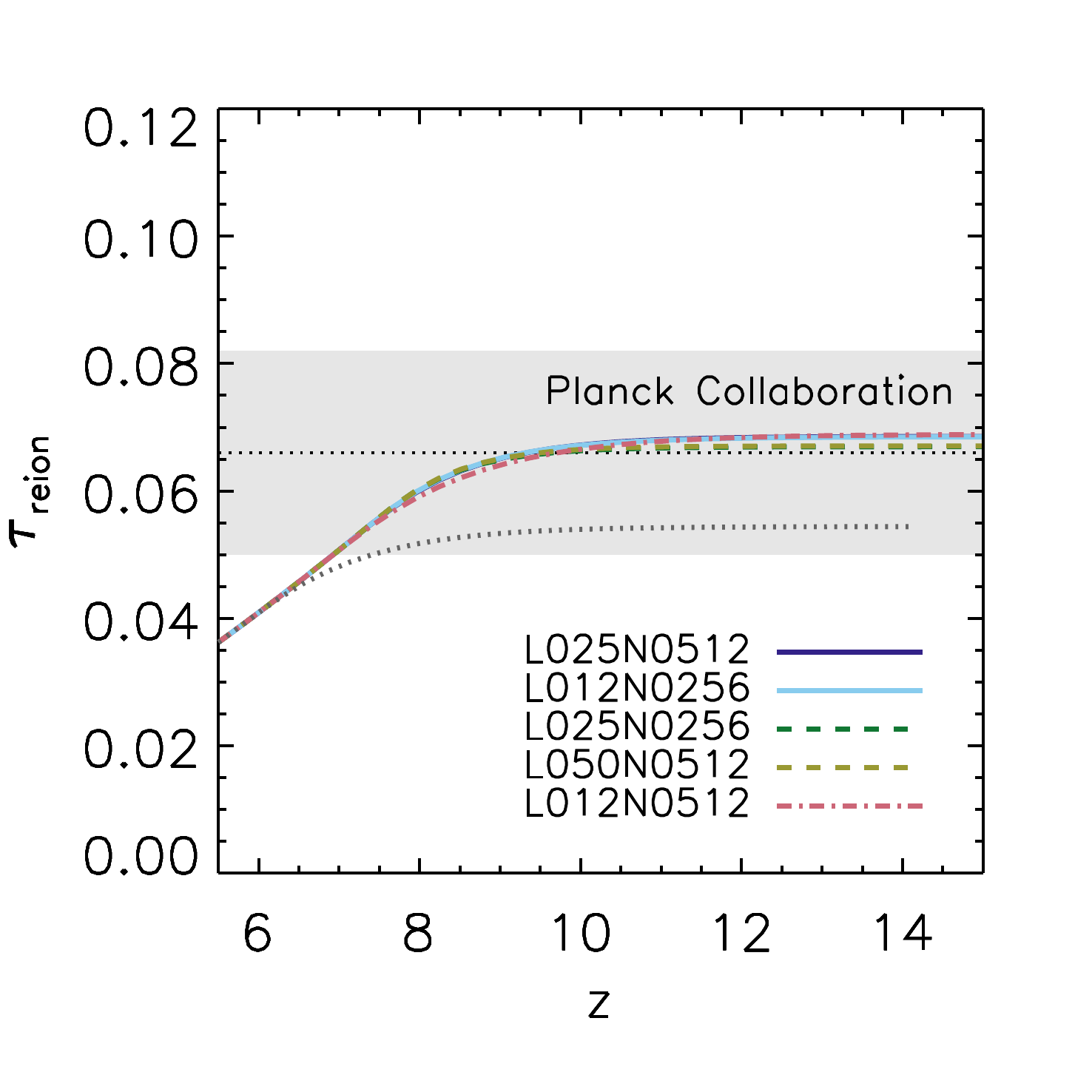}
 
  \end{center}
  \caption{Reionization histories. Different curves correspond to different
    simulations, as indicated in the legend in the right panel. The
    gray dashed curves show a simulation at the resolution of our reference simulation
    but with a smaller subresolution
escape fraction $f_{\rm esc}^{\rm subres} = 0.3$. {\it Left:} mean volume-weighted neutral hydrogen fraction. The
    (model-dependent) constraints on the neutral fraction are taken
    from measurements of QSO damping wings (\protect\citealp{Schroeder2013}),
    Ly-$\alpha$ clustering (\protect\citealp{Ouchi2010}),
    GRB damping wings (\protect\citealp{McQuinn2008}),
    Dark Lyman-$\alpha$ forest pixels (\protect\citealp{McGreer2015}),
    Ly-$\alpha$ emitters (\protect\citealp{Ota2008}, \protect\citealp{Ouchi2010}), 
    Ly-$\alpha$ emission fraction (\protect\citealp{Ono2012},
      \protect\citealp{Robertson2013}, \protect\citealp{Tilvi2014},
      \protect\citealp{Pentericci2014},
      \protect\citealp{Schenker2014}) and the Ly-$\alpha$
      forest (\protect\citealp{Fan2006}), following the discussion of Figure~3 in
    \protect\cite{Robertson2015}. Thanks to
    calibration, all simulations reach a neutral fraction of 0.5
    (dotted line) at
    $z \approx 8.3$, independently of resolution. The
    duration of reionization, here defined as the difference in redshifts
    at which the neutral fraction is between 0.95 and 0.05, increases
    slightly with increasing resolution. {\it Right:}  reionization optical depth.  The simulated reionization
    histories are consistent with the observational constraint
    (dotted line and grey 1-sigma error band; \citealp{Planck2015}).}
  \label{fig2}
\end{figure*}

\subsection{Ionizing luminosities} 
We use the \cite{Schaerer2003} models to
  compute ionizing luminosities for the star particles in our simulations and
  adopt a Chabrier IMF in the range $0.1-100 \Msun$, consistent with
  our implementation of star formation. \cite{Schaerer2003}  does not
  tabulate models for the adopted IMF, but only for a \cite{Salpeter1955} IMF in
  the range $1-100 \Msun$. We therefore correct the
  tabulated luminosities by a factor 0.4 to account for the difference in the mass
  range and by a factor 1.7 to account for the difference in the
  shape between the tabulated and target IMFs. 
\par
  New with respect to PSD15 is that we compute
  metallicity-dependent luminosities, where we use the nearest
  available model metallicity larger than or equal to the metallicity of the
  gas particle from which the star particle was born. We approximately account
  for the pre-enrichment with metals by unresolved minihalos by adopting a
  metallicity floor (only) for the
  computation of the ionizing luminosities, $Z_{\rm min} = 10^{-5} =
  5\times 10^{-4} \Zsun$ (e.g., \citealp{Bromm2011}).
\par
As in PSD15, we multiply the luminosities of star particles by the subresolution
escape fraction $f_{\rm esc}^{\rm subres}$ to account for the removal of
photons due to absorption in the unresolved interstellar medium (ISM) and for 
uncertainties in the intrinsic luminosities of the stars determined by the
population synthesis model. New is
that we calibrate the subresolution escape fraction $f_{\rm esc}^{\rm subres}$  separately
for each simulation. The calibration ensures that all simulations
achieve a volume-weighted hydrogen ionized fraction $x^{\rm V}_{\rm HII} =
0.5$ at nearly the same redshift $z \approx 8.3$ (see
Table~\ref{tab:sims}). Note that the total escape
fraction is the product $f_{\rm esc} = f_{\rm esc}^{\rm subres} f_{\rm
  esc}^{\rm res}$, where $1-f_{\rm  esc}^{\rm res} \ge 0 $ is the fraction
of photons absorbed by the gas particles in the galaxy hosting the
star particles. The latter escape fraction is an outcome of the RT and
is thus not calibrated. The calibration of the subresolution escape fraction will be
discussed in more detail in Section~\ref{sec:calib} below.

\subsection{Ionizing radiative transfer} 
We use the RT code TRAPHIC (\citealp{Pawlik2008}; \citealp{Pawlik2011};
\citealp{Pawlik2013}; \citealp{Raicevic2014}) to transport hydrogen
and helium ionizing photons. TRAPHIC transports radiation directly on the spatially adaptive,
unstructured computational grid traced by the SPH particles and hence exploits
the full dynamic range of the hydrodynamic
simulation. Figure~\ref{fig:zoom} shows that this lets us capture
small-scale structure in the ionized fraction in cosmological volumes.
\par
TRAPHIC further employs
directional averaging of the radiation field to render the computational cost
of the RT independent of the number of sources. This lets us cope efficiently
with the large number of ionizing sources typical of reionization simulations.
\par
Photons are transported in the grey approximation using a single frequency
bin. We compute grey photoionization cross sections  $\langle \sigma_{\rm HI}
  \rangle = 2.93\times10^{-18}\cmsqi$, $\langle \sigma_{\rm HeI} \rangle =
  1.07\times10^{-18}\cmsqi$, $\langle \sigma_{\rm HeII} \rangle =
  1.00\times10^{-21}\cmsqi$ using the fits to the frequency-dependent
  photoionization cross sections by \cite{Verner1996}. The 
  kinetic energy of the electron freed upon photoionizations of an ion of HI, HeI and HeII 
  is $\langle \varepsilon_{\rm HI} \rangle = 3.65 \eV$,
  $\langle \varepsilon_{\rm HeI} \rangle = 4.44 \eV$, and $\langle
  \varepsilon_{\rm HeII} \rangle = 4.10 \eV$. All other parameters of the RT are
  as described in PSD15. In particular, photons leaving the box re-enter the box on the
  opposite side, that is, we are adopting periodic boundary conditions for the radiation. This 
  lets us track the build-up of the UV background in a cosmological setting.

\subsection{Stellar feedback}
We account for the injection of mass and energy by both core-collapse and Type Ia
SNe as well as the mass loss due to winds from massive stars and asymptotic giant branch
(AGB) stars. This gives rise to a collective feedback on the properties of
galaxies to which we refer as stellar feedback. Most of
that feedback is due to the energy released by massive stars.
\par
For each core-collapse SN that occurs, $f_{\rm SN} \times 10^{51} \erg$ of
thermal energy is stochastically injected among a subset of the neighboring
SPH particles, using the method of
\cite{DallaVecchia2012}. The parameter $f_{\rm SN}$ accounts for radiative
energy losses that our simulations may under- or overestimate because
they lack both the resolution and the physics to accurately model the
ISM. Our simulations also lack the resolution to distinguish between
different types of energy feedback, such as SNe and radiatively driven
winds. The factor $f_{\rm SN}$ can hence also be thought to account
for the injection of energy by sources other than SNe. We improve on PSD15 by calibrating $f_{\rm SN}$ separately for each simulation
to bring simulated SFR functions in even better agreement
with each other and with observational constraints. The feedback efficiency
parameter therefore depends on resolution (see Table~\ref{tab:sims}).
The calibration of the stellar feedback will be described below in Section~\ref{sec:calib} in more detail. New is
also that those gas particles that have been heated by core-collapse SNe are
immediately activated for the time integration so that they can react promptly to the
change in energy, as described in \cite{Durier2012}.
\par
Our numerical implementation of the age-dependent injection of energy by Type
Ia SNe follows the deterministic method described in \cite{Schaye2010}. For
each Type Ia SN that occurs, $10^{51} \erg$ of thermal energy is injected
and distributed among all neighboring gas particles. Unlike for core-collapse
SNe, we do not activate neighboring particles
for the prompt response to the change in their energy. The less accurate
treatment of Type Ia SNe is for computational convenience and is justified because they contribute
little to the total energy released by a stellar population above $z \gtrsim
6$, the range of redshifts relevant in this work (e.g., \citealp{Wiersma2009}). In future simulations,
especially when simulating down to lower redshifts, it may make sense to adopt an implementation of Type Ia SNe that is more similar
to that of Type II SNe.
\par
Finally, the implementation of mass loss follows \cite{Wiersma2009}.

\subsection{Identification of galaxies} 
To extract galaxies we post-process the simulations using Subfind (\citealp{Springel2001}; \citealp{Dolag2009}),
  as described in PSD15.

\section{Calibration}
\label{sec:calib}
We calibrate the two main parameters of our simulations, the stellar 
feedback efficiency $f_{\rm SN}$ and the subresolution escape fraction of 
ionizing radiation $f_{\rm esc}^{\rm subres}$, to yield agreement between 
simulated and observed SFR functions at $z = 7$ independently of the resolution and to achieve a 
reionization redshift that is independent of the resolution and box size and in agreement with 
observational constraints on the optical depth for electron scattering
from measurements of
the CMB. The reionization redshift is thereby defined as the redshift
at which the mean volume-weighted neutral hydrogen fraction reaches
0.5. Calibration is required, because cosmological simulations
like ours lack both the physics and the
resolution to predict SFRs and reionization histories from first
principles (e.g., \citealp{Rahmati2013b}). A similar calibration was carried out in \cite{Gnedin2014}
and \cite{Schaye2015}.
\par
Each of these two numerical parameters can impact both the SFR functions and the redshift 
of reionization. This could potentially complicate the calibration, which might need to proceed
iteratively, calibrating each parameter at a time, possibly converging
slowly. However, reionization does not significantly impact the SFR
function at the SFRs
at which it is observationally constrained, it only affects the SFR
function at much lower
SFRs. We can therefore adopt the 
following two-step calibration procedure. First, we choose for each simulation
a stellar feedback efficiency that yields agreement between observed and simulated
SFRs. Second, we choose for each simulation a subresolution escape fraction such 
that the redshift of reionization, defined as the redshift at which half of
the simulation volume is ionized, equals $z = 8.3$. The latter implies  
an optical depth to electron scattering in agreement with constraints
from observations of the CMB ($\tau = 0.066 \pm 0.016$; \citealp{Planck2015}).
\par
The calibrated values of the feedback efficiency and the subresolution
escape fraction will generally depend on box size and resolution, for several reasons. First,
the radiative losses that are simulated explicitly may change if we
better resolve the structure of the ISM. Keeping the feedback
efficiency parameter fixed while changing the resolution may therefore
affect the match between
simulated and observed SFR function. Second, the reionization redshift is sensitive
to resolution because resolution impacts the absorption of photons in the
simulation, and it is sensitive to box size unless the box is
much larger than the characteristic size of individual ionized
regions. Third, a change in resolution impacts the SFR function
at small, currently unobservable SFRs, at which the SFR function is not fixed
by calibration. This, in turn, affects the redshift of reionization,
also requiring us to adjust the subresolution escape fraction.
\par
In the following, we discuss the calibration of each of the
parameters in more detail. In addition to the calibrated simulations,
we have carried out simulations in which we have varied the stellar
feedback efficiency and the subresolution escape fraction away from
their calibrated values, including simulations in which we have turned
off SN explosions and photoheating altogether. While this work focuses
on the calibrated set of simulations, these additional simulations can
be used to gain further insight into the role of these parameters and the roles
of galactic winds and photoheating, similar to our approach in PSD15.

\begin{figure*}
   \begin{center}
     \includegraphics[width=0.28\textwidth,clip=true, trim=0 0 0 0,
      keepaspectratio=true]{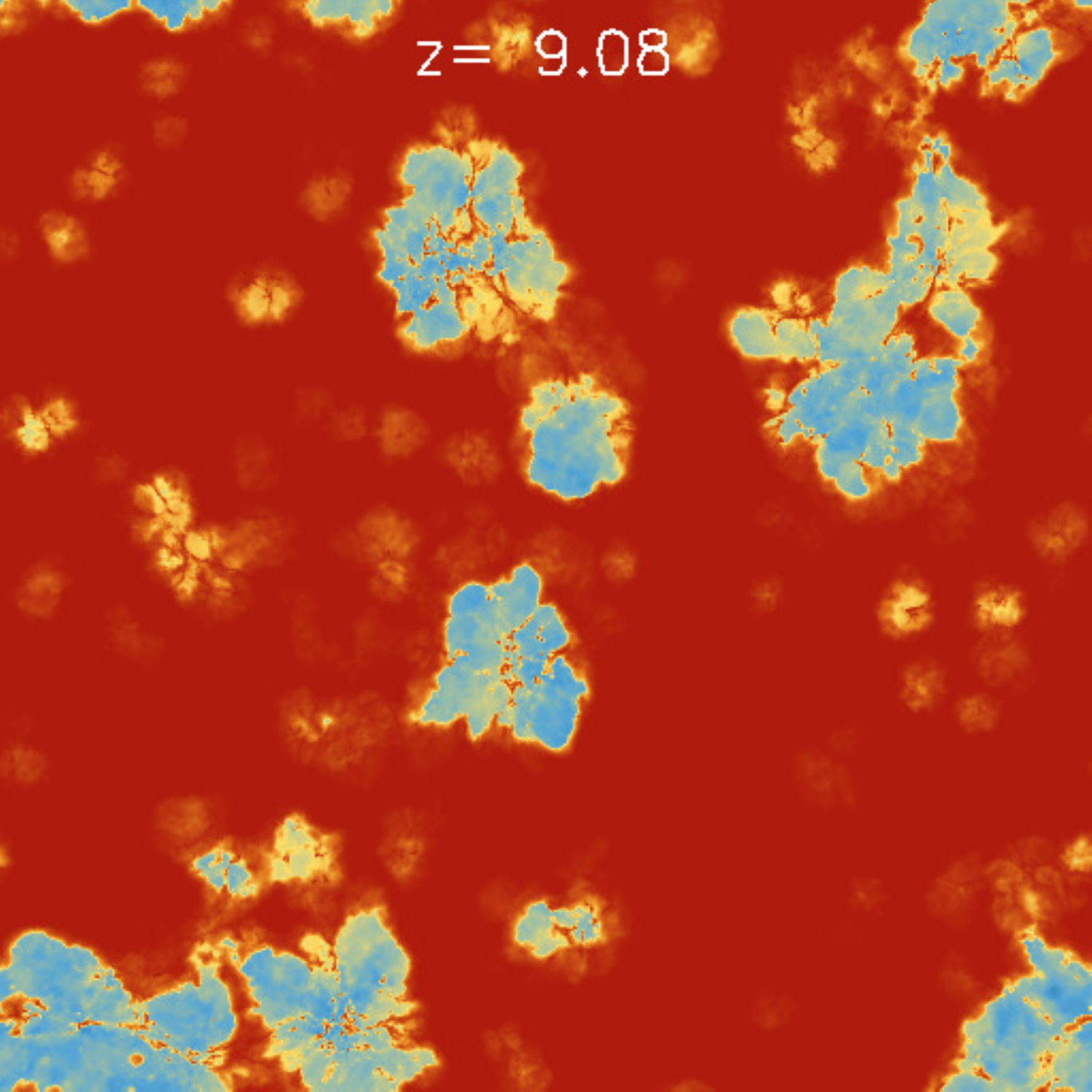}
    \includegraphics[width=0.28\textwidth,clip=true, trim=0 0 0 0,
      keepaspectratio=true]{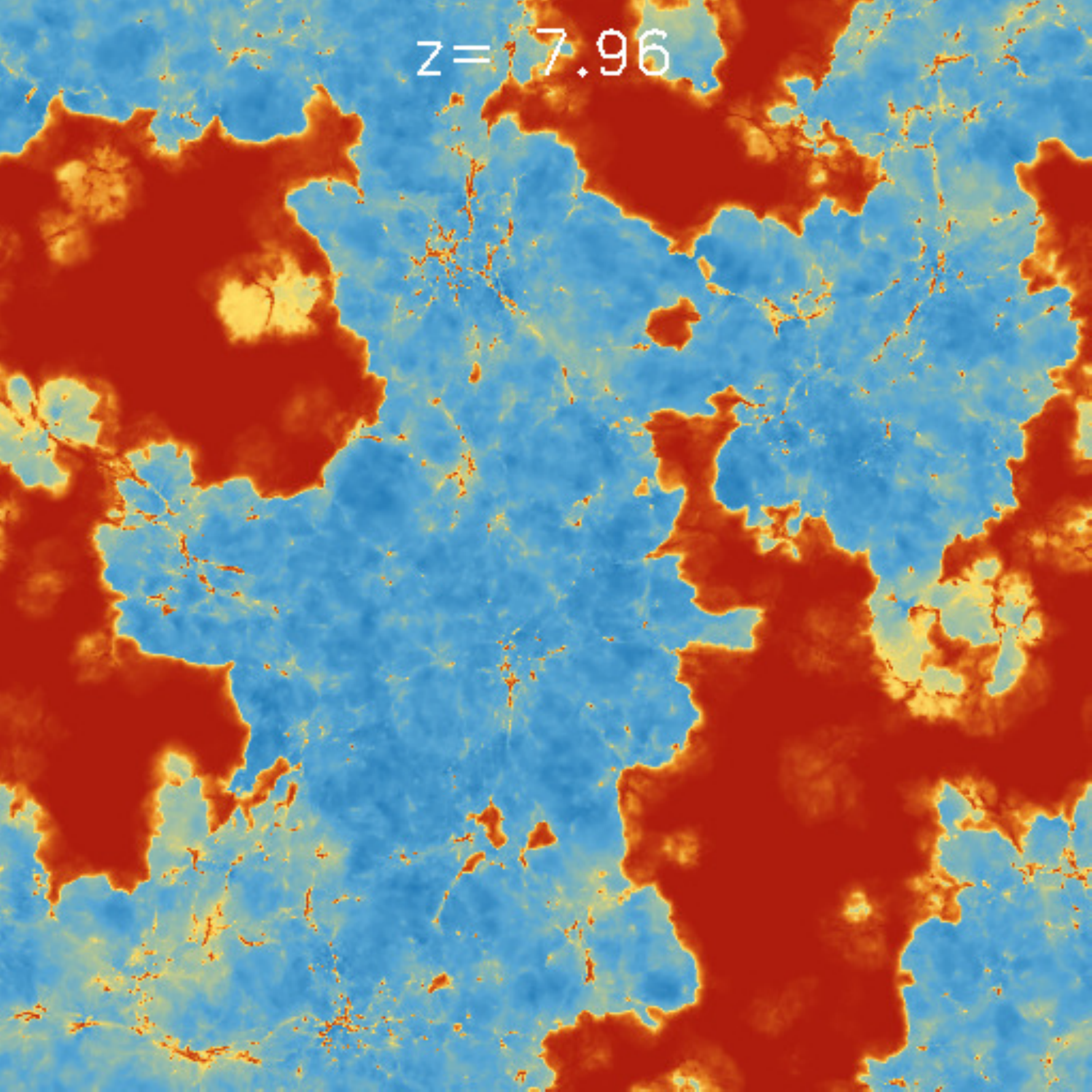}
    \includegraphics[width=0.28\textwidth,clip=true, trim=0 0 0 0,
      keepaspectratio=true]{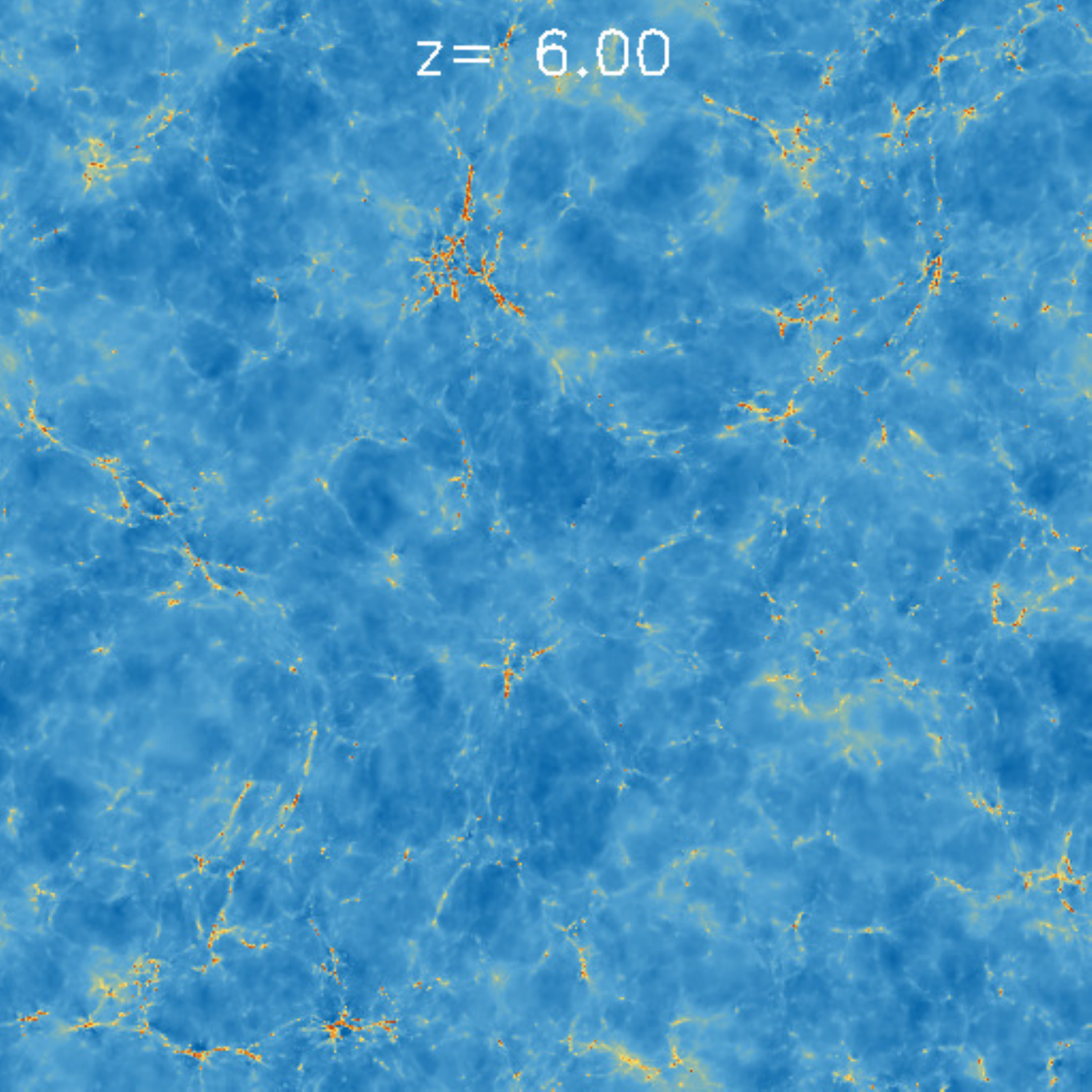}
   \includegraphics[width=0.1\textwidth,clip=true, trim=20 0 0 0,
      keepaspectratio=true]{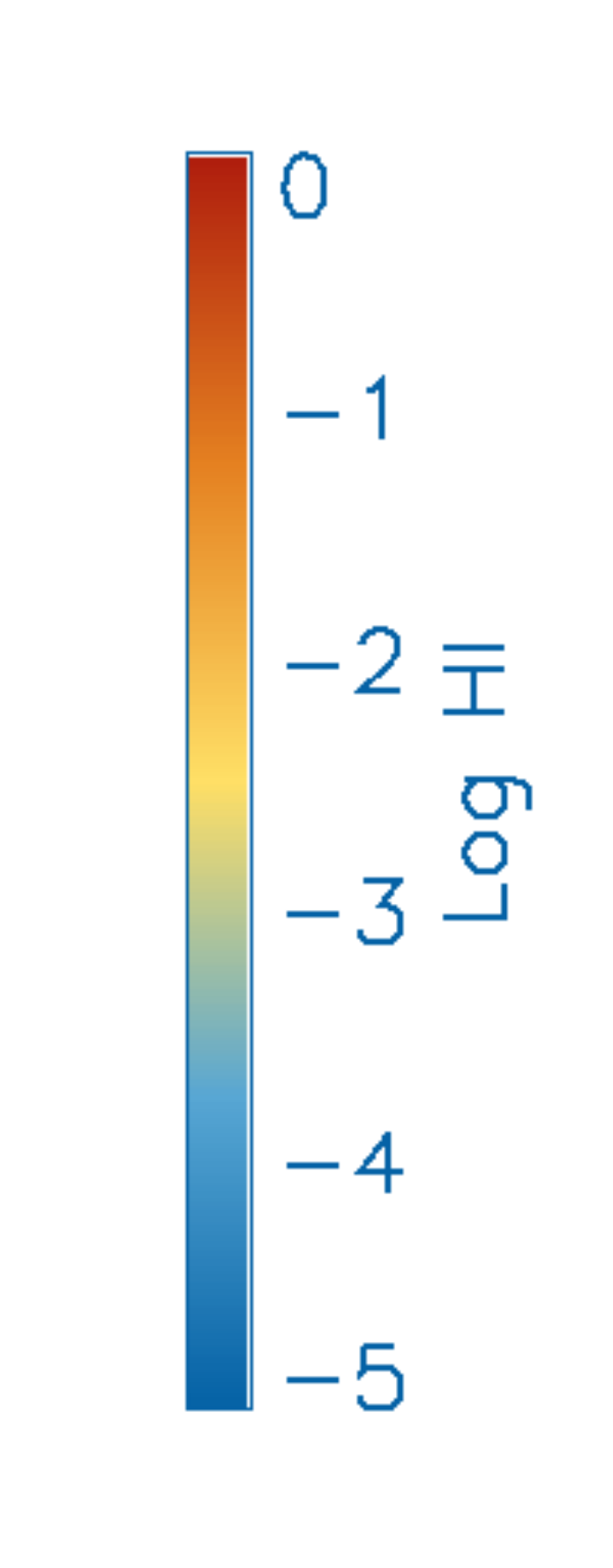}
\\
    \includegraphics[width=0.28\textwidth,clip=true, trim=0 0 0 0,
      keepaspectratio=true]{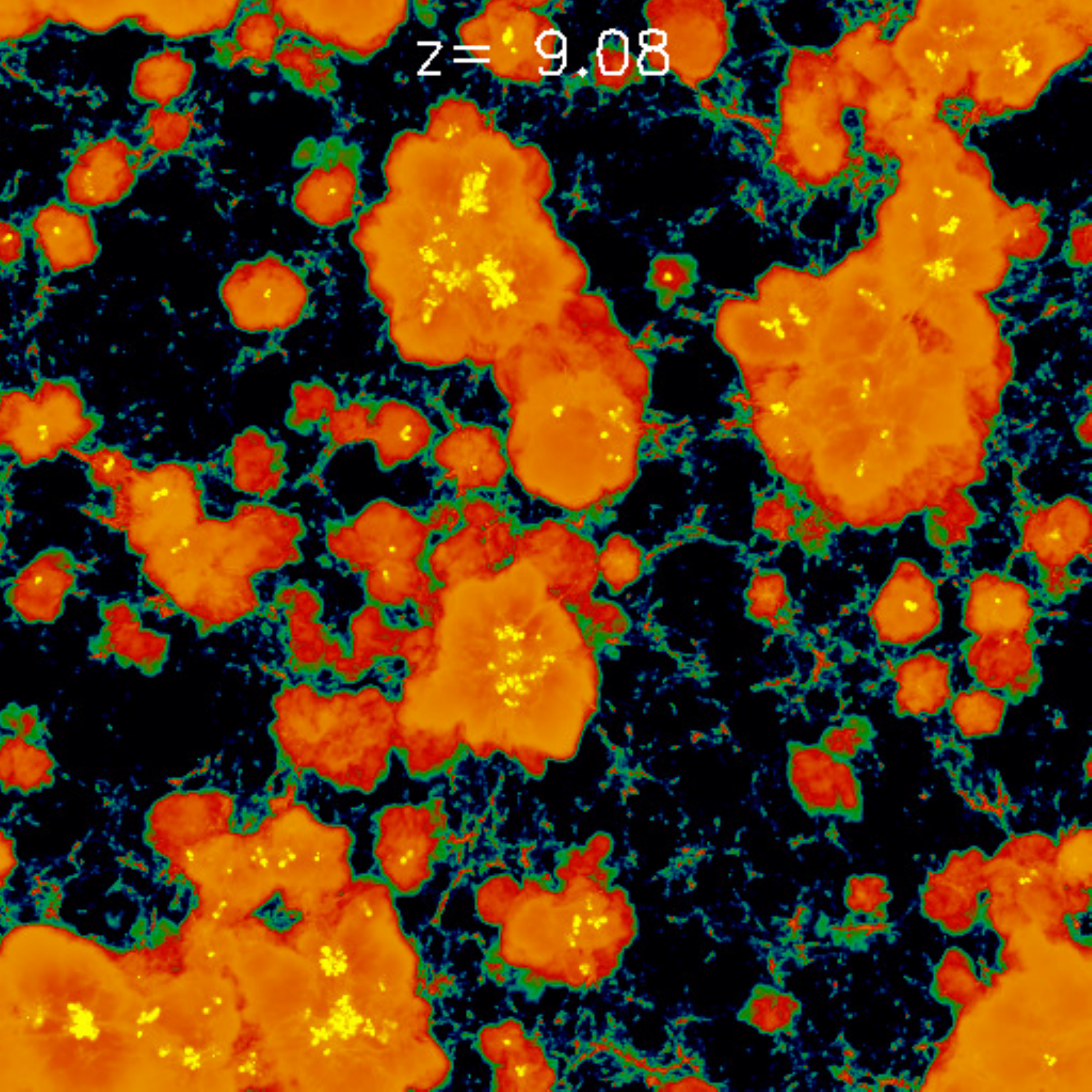}
    \includegraphics[width=0.28\textwidth,clip=true, trim=0 0 0 0,
      keepaspectratio=true]{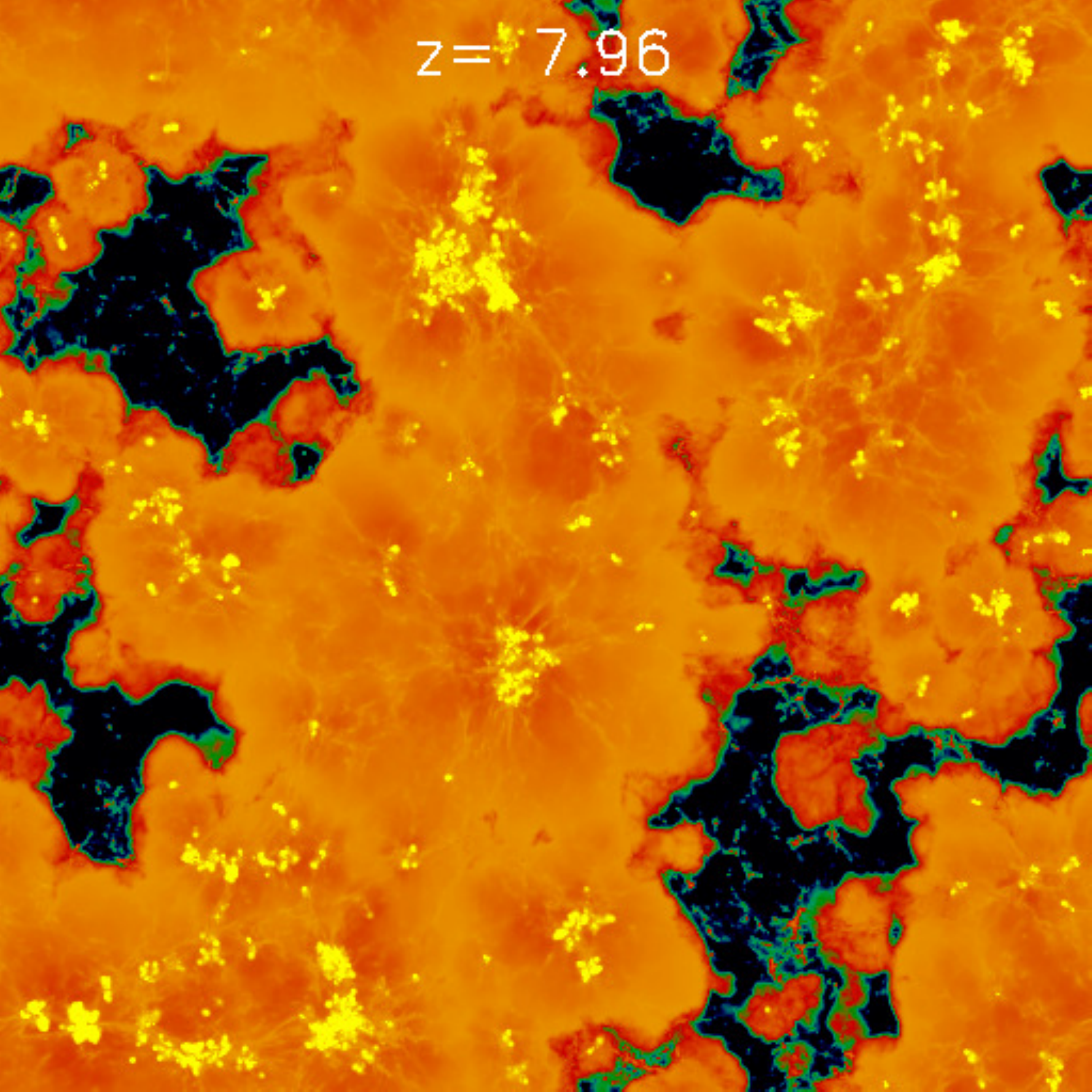}
    \includegraphics[width=0.28\textwidth,clip=true, trim=0 0 0 0,
      keepaspectratio=true]{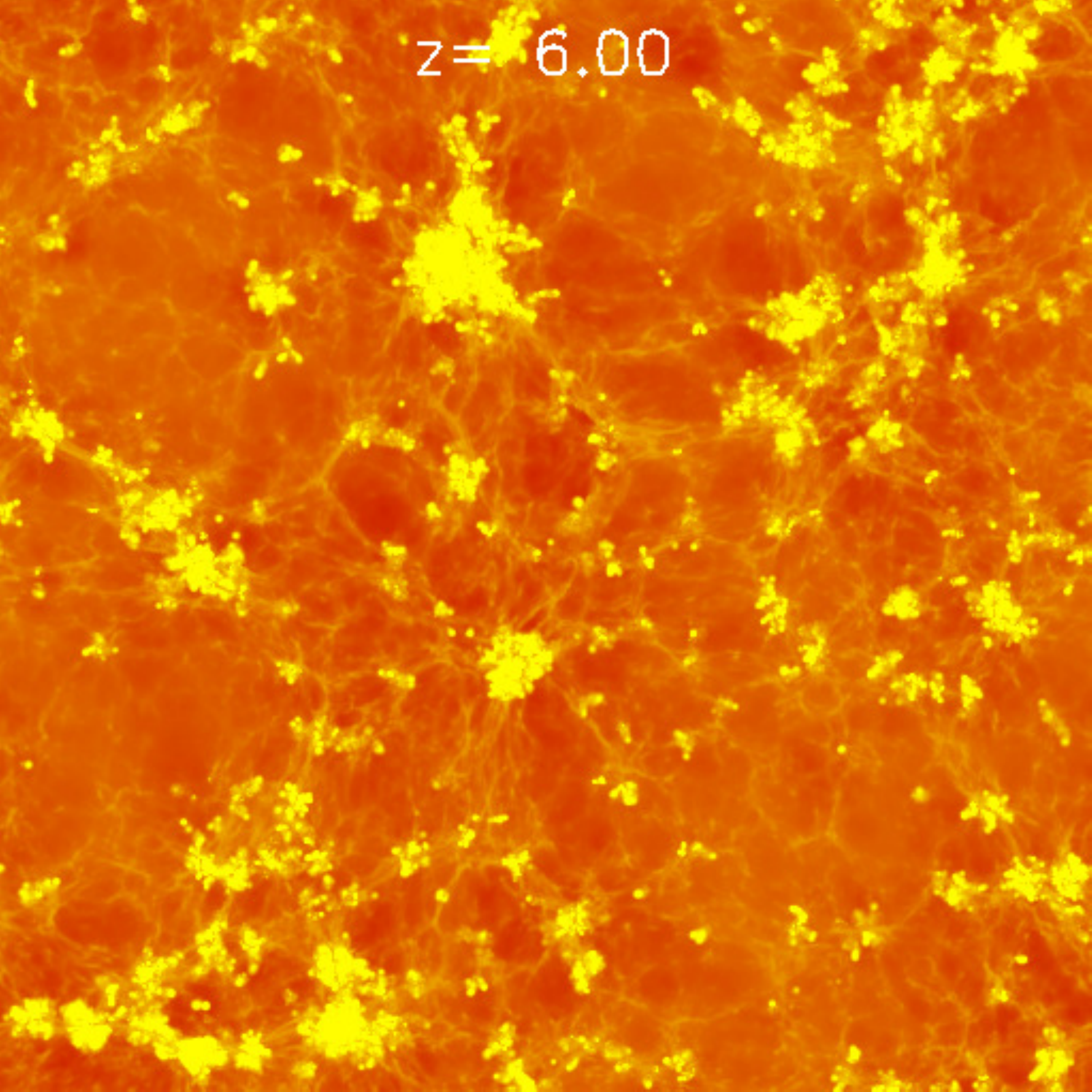}
   \includegraphics[width=0.1\textwidth,clip=true, trim=20 0 0 0,
      keepaspectratio=true]{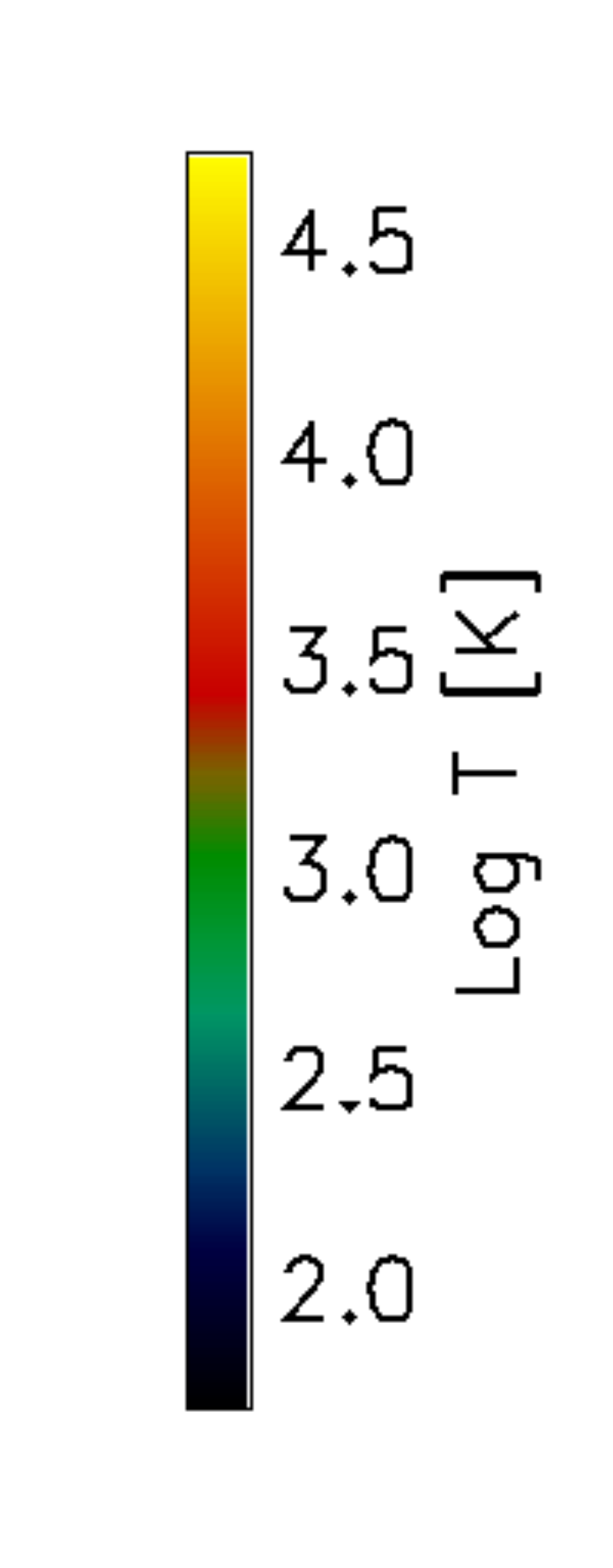}
  \\
    \includegraphics[width=0.28\textwidth,clip=true, trim=0 0 0 0,
      keepaspectratio=true]{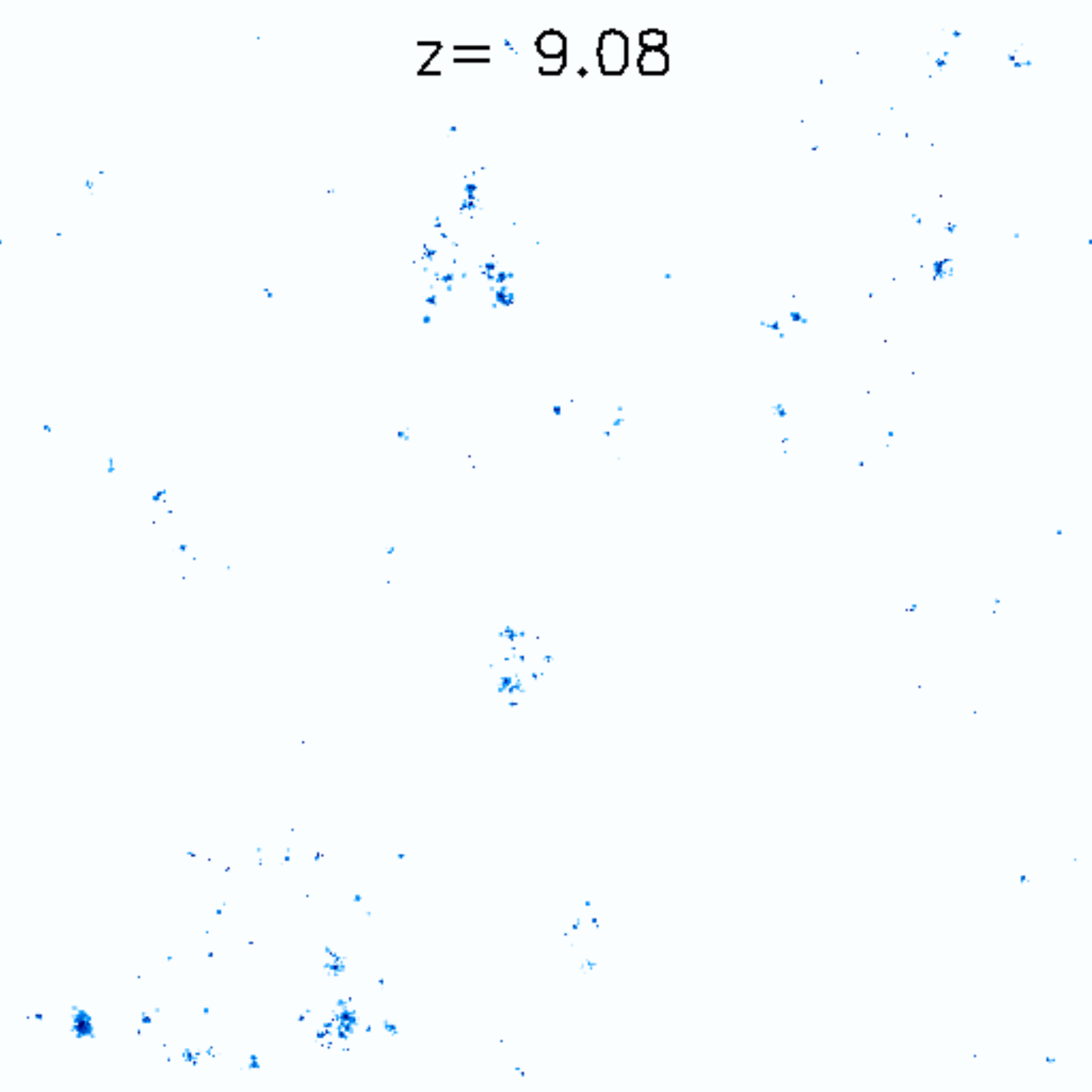}
    \includegraphics[width=0.28\textwidth,clip=true, trim=0 0 0 0,
      keepaspectratio=true]{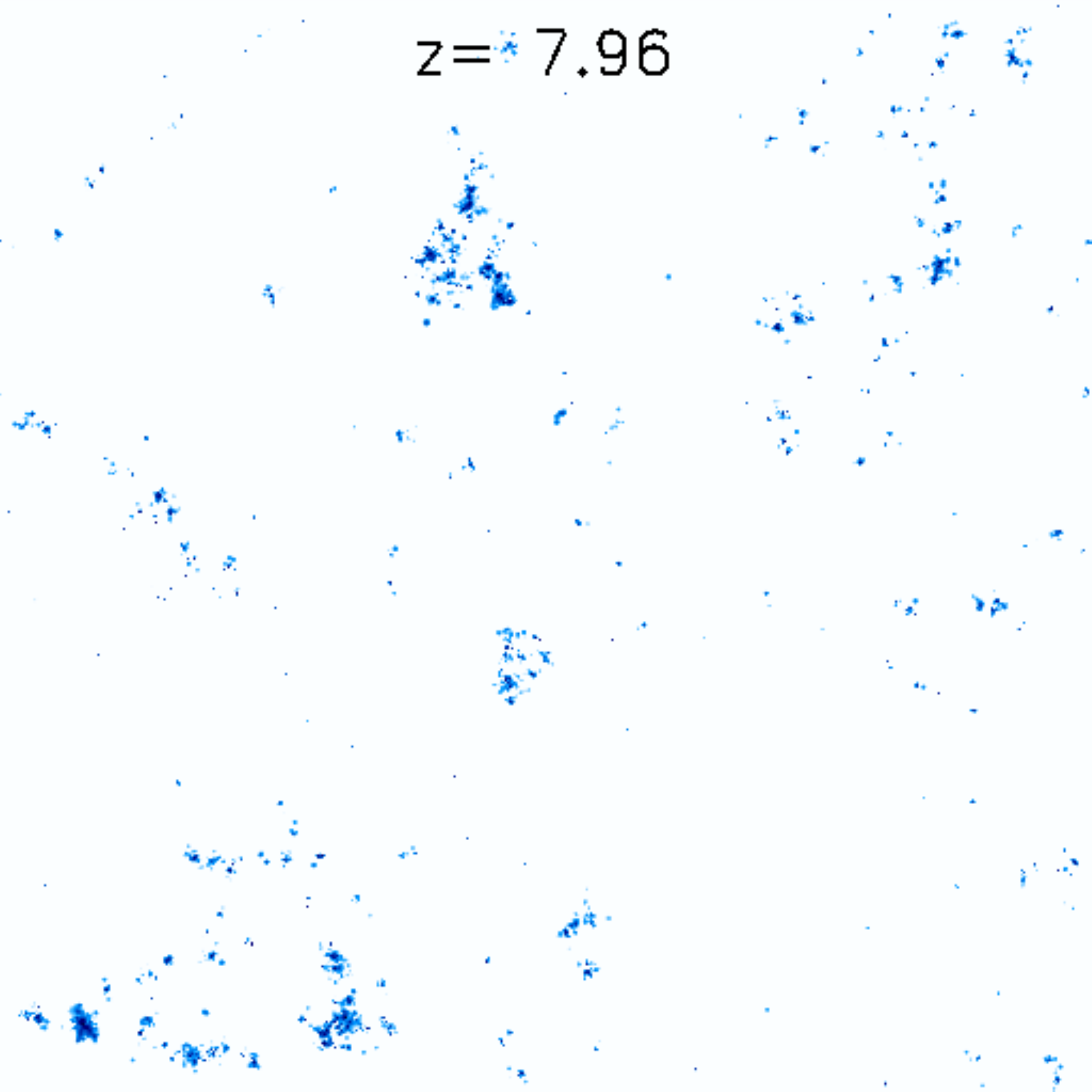}
    \includegraphics[width=0.28\textwidth,clip=true, trim=0 0 0 0,
      keepaspectratio=true]{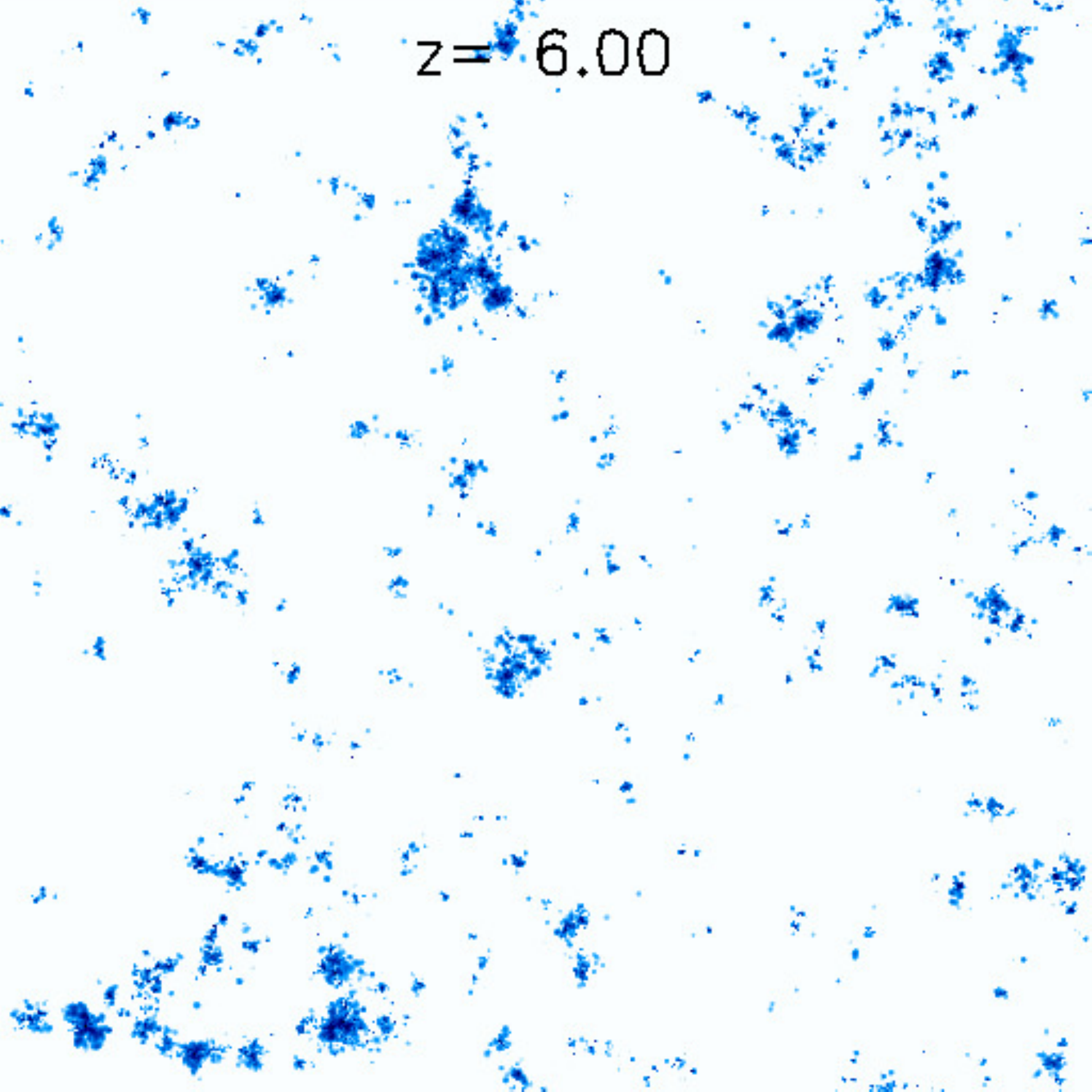}
   \includegraphics[width=0.1\textwidth,clip=true, trim=20 0 0 0,
      keepaspectratio=true]{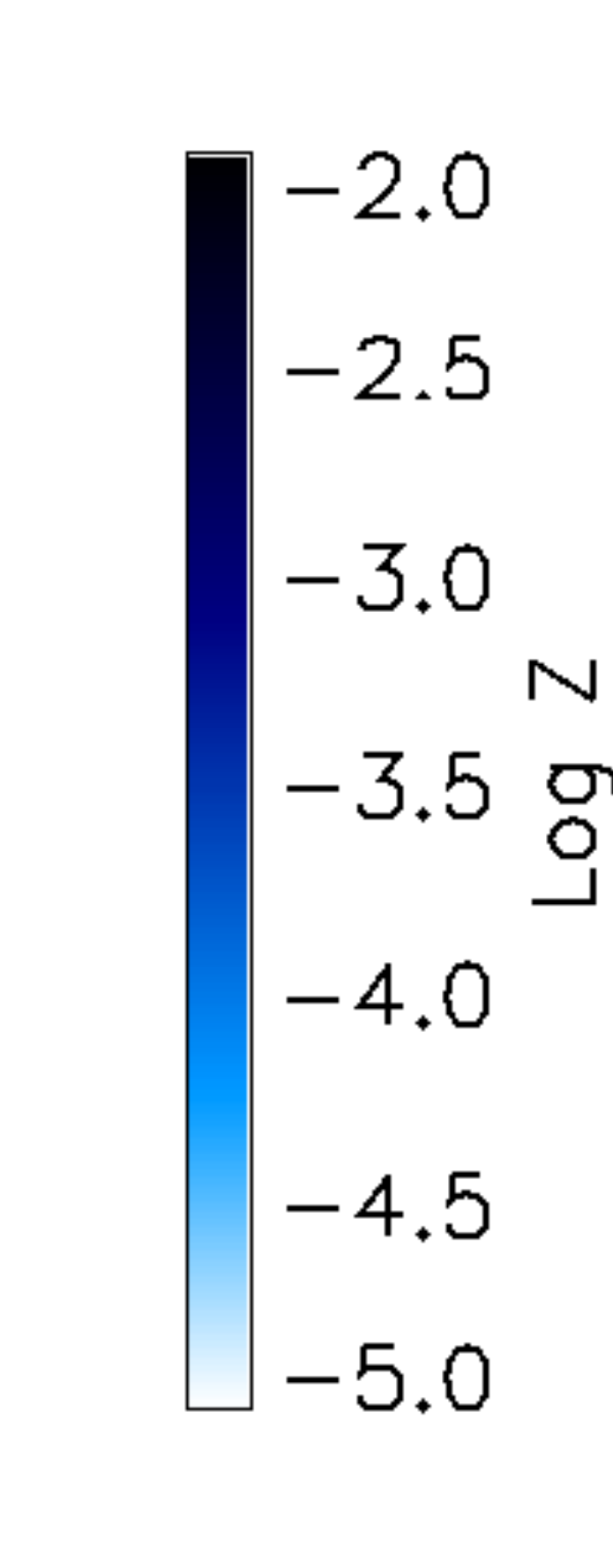}
  \end{center}
  \caption{Snapshots of the reference simulation L025N0512 showing the neutral hydrogen fraction (top row), the
    temperature (middle row), and the gas metal mass fraction (bottom row) at
    three characteristic redshifts $z \approx 9$, 8, and 6, corresponding to 
    the early, middle and post-overlap phases of reionization,
    respectively. Each panel shows a thin central slice through the
    simulated cubical volume. Photoionization heats the
  gas to a few $10^4\K$, and our radiation-hydrodynamical simulations
  let us accurately account for the feedback associated with this
  process. The localized regions of very high temperatures are caused
  by energetic feedback from massive stars, which also enrich the gas
  with metals synthesized in the stars.}
  \label{fig0}
\end{figure*}

\begin{figure}
  \begin{center}
    \includegraphics[width=0.49\textwidth,clip=true, trim=-10 20 20 0,
      keepaspectratio=true]{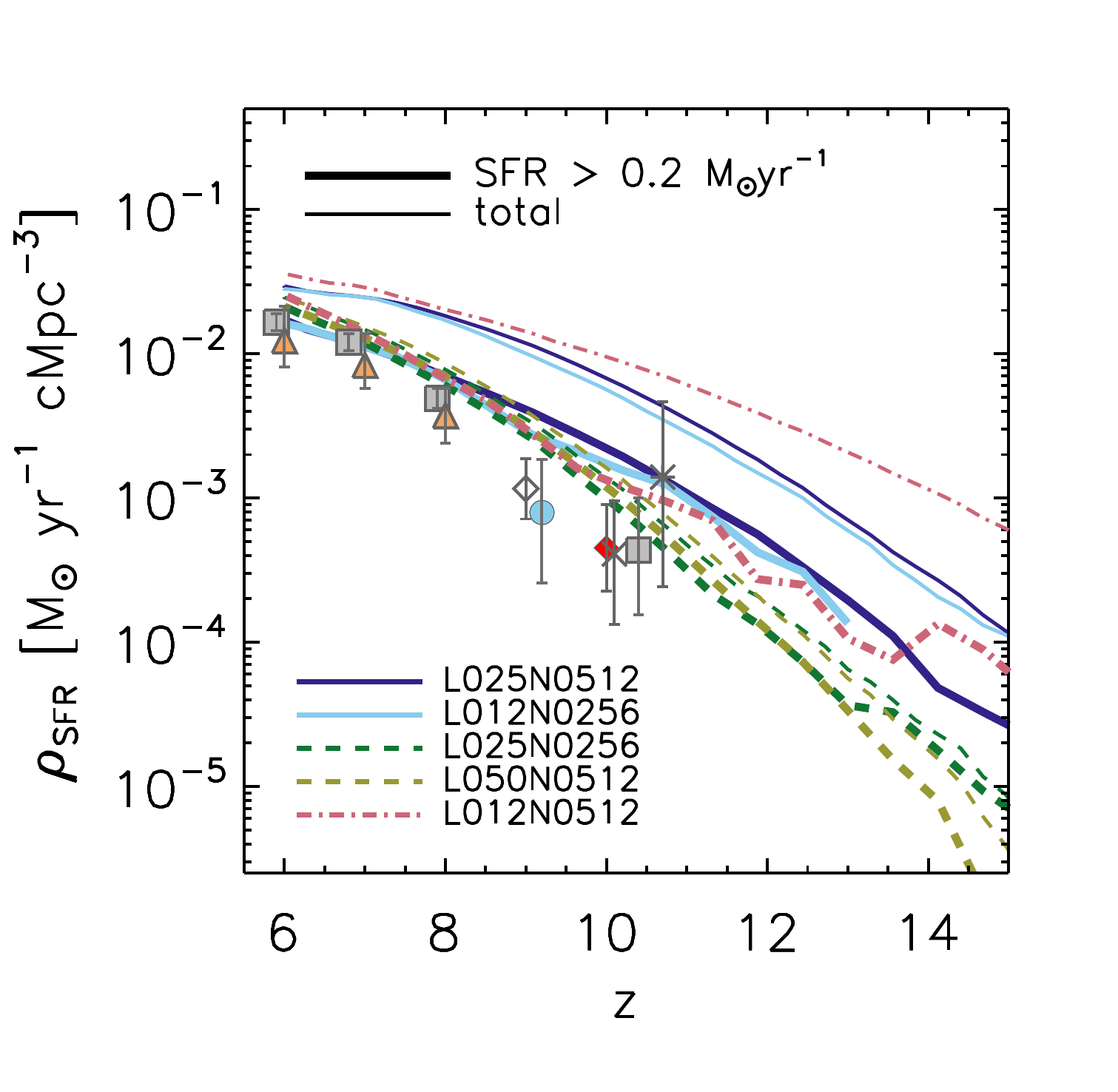}

  \end{center}
  \caption{Effect of box size and resolution on the evolution of the comoving SFR
    density. Curves correspond to different
    simulations as indicated in the legend. Thin and thick curves
    show the total SFR density and the SFR density including only the
    contribution from galaxies with SFRs greater than $0.2 \Msun\invyr$,
    approximately the current detection limit (assuming a Chabrier
    IMF, corresponding to a limiting UV magnitude $M_{AB}  \approx −17$,
    see Figure~\ref{fig1}). Where necessary, we have converted the observed
    SFRs to the Chabrier IMF. Observational data is taken from \protect\cite{Bouwens2015},
    \protect\citet[filled circle]{Bouwens2014},
    \protect\citet[triangles]{Finkelstein2014}, \protect\citet[open diamond]{Oesch2013},
    \protect\citet[filled diamond]{Oesch2014}, \protect\citet[asterisk]{Coe2013}, and
    \protect\citet[cross; revised down by a factor of two following
      \citealp{Oesch2013}]{Ellis2013}. All simulations yield similar SFR densities
    from galaxies above the detection limit at $z \lesssim 10$, 
    independently of the box size and resolution, and in agreement
    with observational constraints.}
  \label{fig1a}
\end{figure}

\subsection{Stellar feedback efficiency}
Figure~\ref{fig1} shows the SFR functions at redshifts $z \approx 6$,
7, 8, and 9. We calibrated the stellar
feedback efficiency $f_{\rm SN}$ at $z = 7$ such that all simulations yield nearly
identical SFR functions when considering galaxies with SFRs
that are accessible by current observations using the HST, i.e., galaxies with
SFRs $\gtrsim 0.2 \Msun \invyr$. Where necessary, before carrying out the calibration, we have divided the observed SFRs, which
assume a Salpeter IMF, by 1.7 to enable comparisons with our simulations, which assume a Chabrier IMF.
\par
Figure~\ref{fig1} demonstrates generally good agreement between simulated and
observationally inferred SFR functions above z = 6. By construction, 
our results are very close to the observational constraints by \cite{Smit2012}
at z = 7, which we used for our
calibration. However, differences with respect to more recent estimates of the SFR
function and at other redshifts are minor. 
\par
The good match was facilitated by our
calibration of the stellar feedback efficiency parameter, since 
stellar feedback dominates over feedback from photoheating in the range of SFRs at
which the SFR function has been observationally inferred
(PSD15). Achieving this match required us to reduce the stellar
feedback efficiency parameter from $f_{\rm SN} = 1.0$ at the lowest
resolution to  $f_{\rm SN} = 0.6$ at the highest resolution  (see
Table~\ref{tab:sims}). This implies that the cooling losses are
smaller at higher resolution, which is consistent with \cite{DallaVecchia2012}. The calibration does not depend
on box size because for our set of simulations, box size impacts
the SFR functions only at very large SFRs at which stellar feedback is inefficient. 
At these large SFRs, additional processes such as, e.g., dust corrections
and black hole feedback may need to be invoked to
reduce any remaining mismatch with the observations. 

\subsection{Escape fraction}
The left panel of Figure~\ref{fig2} shows the evolution of the mean volume-weighted hydrogen
neutral fraction. Because of our calibration of the subresolution
escape fraction,  $f_{\rm esc}^{\rm subres}$, all simulations reach a
neutral hydrogen fraction of 0.5 at $z \approx 8.3$,
independently of the resolution (except for the dotted curve which
shows a simulation with a different calibration using a lower
subresolution escape fraction of 0.3, which is explained in
Section~\ref{Sec:reionization}). Furthermore, as the right panel of
Figure~\ref{fig2} shows, all simulated 
reionization histories are in excellent
agreement with the observational estimate of the electron scattering optical
depth (\citealp{Planck2015}).
\par
Achieving this agreement required us to decrease the subresolution escape fraction
with increasing resolution, from   $f_{\rm esc}^{\rm subres} = 1.3$ at
the lowest resolution to $f_{\rm esc}^{\rm subres} = 0.5$ at the
highest resolution (see Table~\ref{tab:sims}). This points at an increased contribution of
low-mass galaxies as well as a more leaky gas distribution,
facilitating the escape of photons, at increased
resolution. The former can be deducted from the increasing star formation rate density with increasing resolution (see Figure~\ref{fig1a}). However, box size does not impact the choice of escape fraction, because in our simulations, the reionization histories do not change with box size above $\gtrsim 10 \cMpch$ (please see Section 3.1 in PSD15 for a more detailed discussion).

\par
Simulations of reionization can also be calibrated against the optical
depth (e.g., \citealp{Ciardi2012}), but that
does not necessarily guarantee the same redshift of
reionization. Direct calibration against the redshift of reionization,
as done here, still results in an excellent match between simulated
and observationally inferred optical depths, while also rendering
feedback from reionization insensitive to changes in resolution and box
size. Because of this, a fixed redshift of reionization facilitates
the analysis of the impact of resolution and box size on our
simulations.
\par
Figure~\ref{fig2} shows that in our calibrated set of simulations,
reionization occurs earlier than inferred from observations of Lyman-$\alpha$
emitters and quasar damping wings at $z\approx 7$. However, as we
discuss in Section~\ref{Sec:reionization} below, 
the observational constraints are strongly model-dependent, and therefore
matching these constraints was not our primary
aim. Instead, the goal of the calibration was to yield a
reionization redshift that is approximately independent of
resolution, that implies an electron optical depth consistent with
observations, and that is sufficiently high such that we can investigate
the impact of the feedback from photoheating at redshifts $z \gtrsim
6$. 
\par
The subresolution escape fraction merely multiplies the stellar
luminosities and thus does not fully specify the escape fraction of
galaxies in our simulations. The fraction of ionizing photons that escape a galaxy and
are available to reionize the IGM is the product $f_{\rm esc} = f_{\rm
esc}^{\rm subres} f_{\rm esc}^{\rm res}$, where $1-f_{\rm esc}^{\rm
res} \ge 0 $ is the fraction of photons absorbed by simulation gas particles on
their way to the IGM. Because the subresolution escape fraction is not
a physical quantity, we are free to set it to values larger than
unity, as we did in our low resolution
simulations, as long as the number of ionizing photons reaching the
IGM remains consistent with expectations from population synthesis
models. A value $ f_{\rm
esc}^{\rm subres} >1 $ then implies that the RT simulation
overestimates the fraction of photons that are absorbed in the ISM,
for example because it does not resolve holes in the ISM, or that we
need to compensate for missing radiation from unresolved low-mass
galaxies. Finally, we note that the calibrated subresolution escape
fraction is chosen independently of the properties of the galaxies and time,
while in reality the escape fraction may differ strongly between galaxies and change with time
(e.g., \citealp{Wise2009}; \citealp{Paardekooper2015}).
\par

\section{Results}
Here we discuss results from a first analysis of the simulations in
Table~\ref{tab:sims}. Because simulation L100N1024 was stopped
at $z = 8.4$, we will show the SFR function at  $z = 9$ derived from this
simulation, but otherwise omit it from our analysis.  

\subsection{Overview}
Figure~\ref{fig0} shows images of the neutral hydrogen fractions, gas
temperatures and gas metallicities in our reference simulation L025N0512 at
three characteristic redshifts $z \approx 9$, 8, and 6, corresponding to
volume-weighted ionized fractions of $x_{\rm HII} \approx 0.3$, 0.6, and
1, respectively. Stellar ionizing radiation creates individual ionized regions that
grow and overlap. While the mean ionized fraction 
of these regions is very high, the neutral
fraction can locally reach high values as gas is shadowed or self-shields from
the ionizing radiation (see also Figure~\ref{fig:zoom}). 
\par
The reionization of the gas is accompanied by an increase in its temperature
to about $10^4 \K$, the characteristic temperature of gas photoionized by
stellar radiation. Locally the gas reaches higher temperatures, up to $10^7\K$, where
it is heated by structure formation shocks and explosive stellar
feedback. Galactic winds enrich the gas with metals synthesized in the stars. 
\par
Photoheating raises the pressure in the galaxies and the surrounding reionized
regions, which impedes the gravitational collapse of the gas in galaxies and
in the IGM. As a consequence, reionization is associated with a
decrease in the rates at which stars are formed and thus provides a negative
feedback on reionization (e.g., \citealp{Shapiro1994};
\citealp{Gnedin2000}; \citealp{Pawlik2009}; \citealp{Noh2014}; PSD15). In the IGM, 
however, the dynamical impact from photoheating reduces the recombination
rate, and this provides a positive feedback on reionization (e.g., 
\citealp{Wise2005}; \citealp{Pawlikclump2009}; \citealp{Sobacchi2014}).  Both types of
feedback, negative and positive, are captured self-consistently by our
radiation-hydrodynamical simulations.
\par
In the following, we will provide a more quantitative discussion of
our simulations, and investigate the dependence on box size and resolution.

\begin{figure*}
  \begin{center}

    \includegraphics[width=0.49\textwidth,clip=true, trim=0 0 0 0,
      keepaspectratio=true]{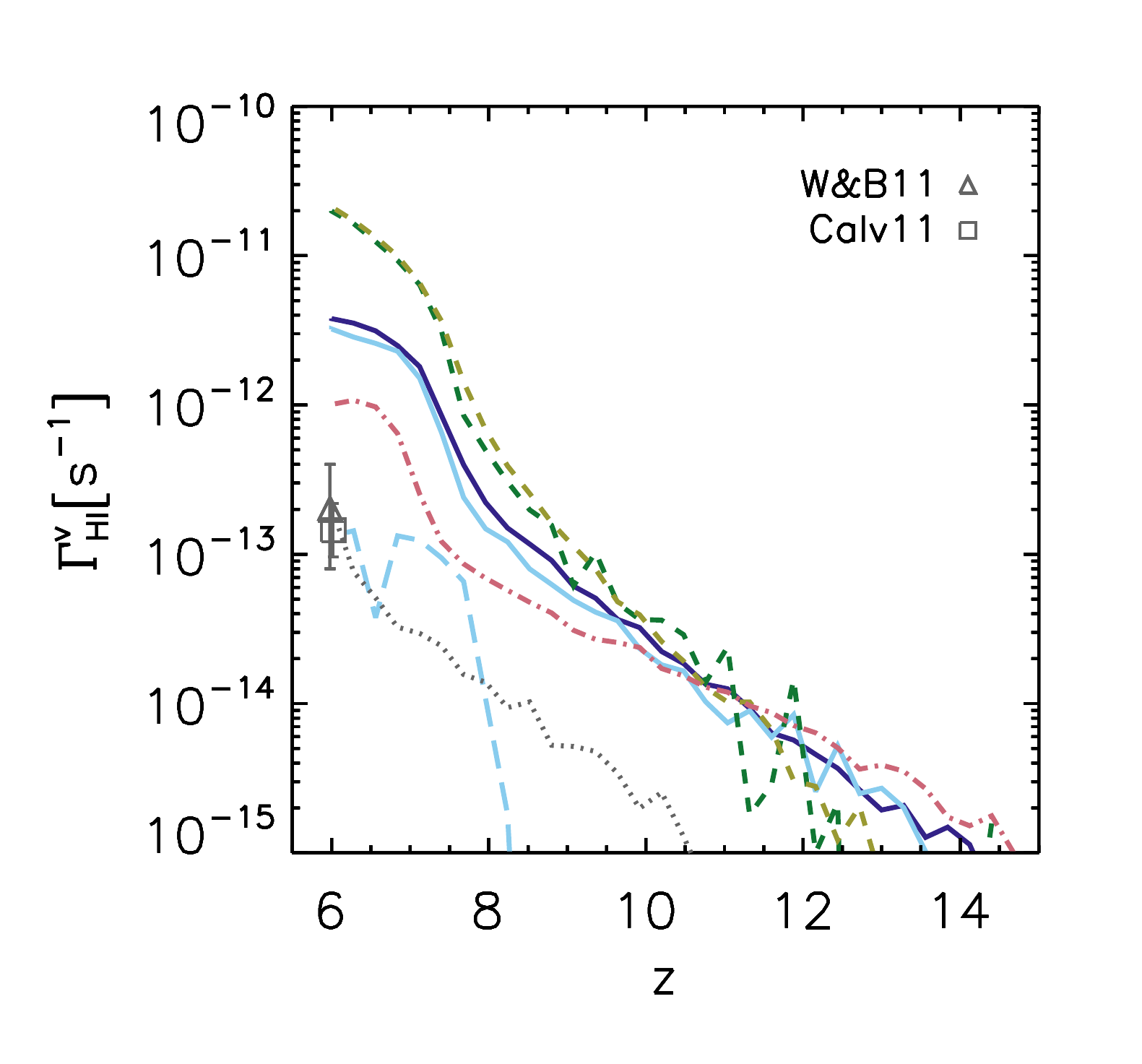}
    \includegraphics[width=0.49\textwidth,clip=true, trim= 0 0 0 0,
      keepaspectratio=true]{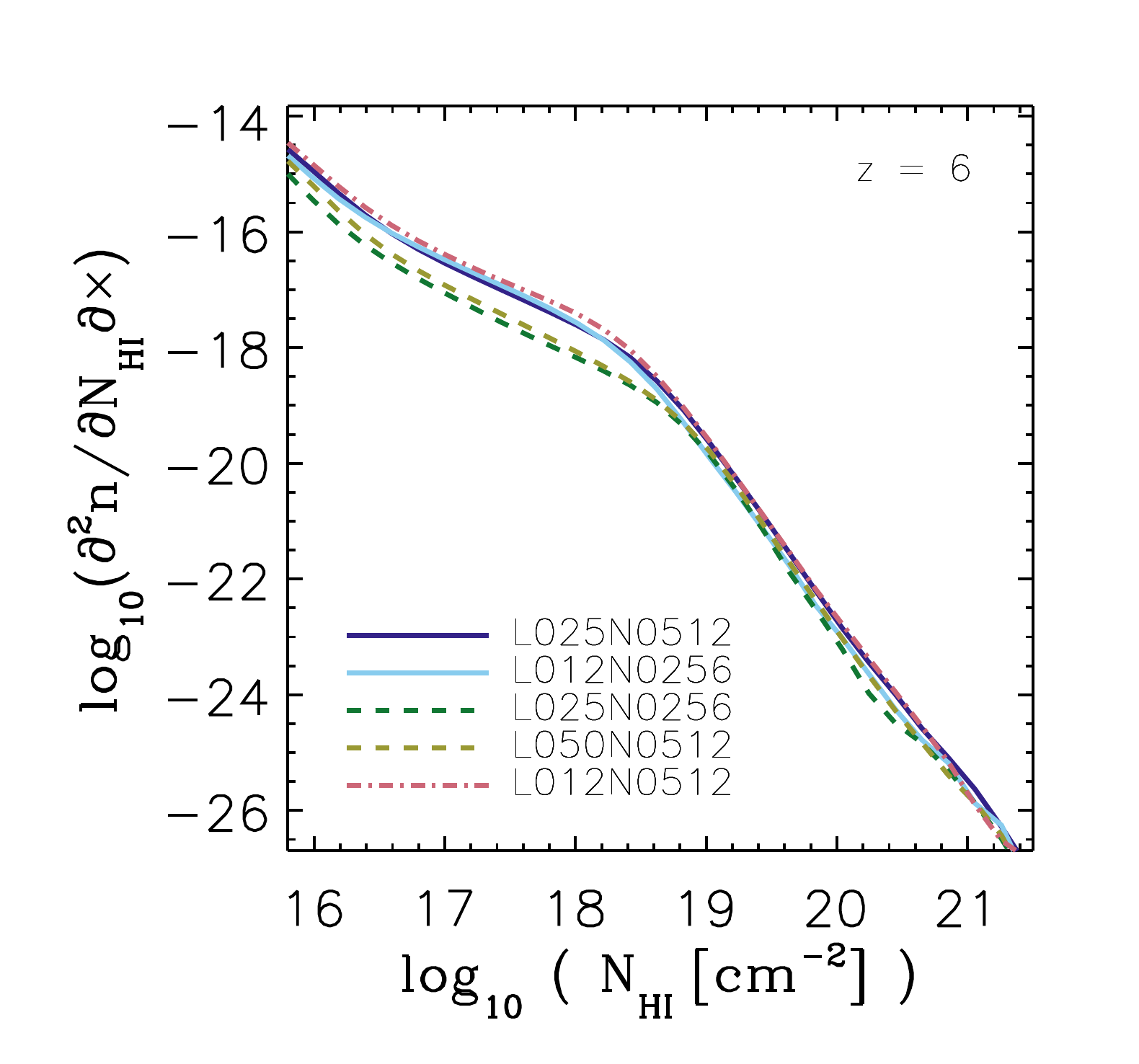}

  \end{center}
\caption{{\it Left}: Mean volume-weighted hydrogen photoionization rate. At $z < 10$, the simulations reveal a marked trend of decreasing photoionization rate with increasing resolution. The observational estimates are from \citet[triangle]{Wyithe2011} and \citet[square]{Calverley2011}. The dotted curve shows a simulation at the resolution of our reference simulation but with a smaller subresolution escape fraction $f_{\rm esc}^{\rm subres} = 0.3$ (see Figure~\ref{fig2}). Moreover, the light-blue long-dashed curve shows the median photoionization rate at the IGM densities for the L012N0256 simulation. The significant qualitative and quantitative differences between the mean and median photoionization rates indicate a large scatter in the distribution of photoionization rates in the simulations. {\it Right: } Neutral hydrogen column density distribution function at $z = 6$. The abundance of Lyman-limit systems, i.e., absorption systems with neutral hydrogen column densities $N_{\rm HI}\gtrsim 10^{17} \cmsi$ increases systematically with increasing resolution. }

\label{fig2a}
\end{figure*}

\subsection{Star formation history}
Figure~\ref{fig1} shows that thanks in part to the calibration of stellar
feedback near $z = 7$, the simulated and
observed SFR functions at $z \gtrsim 6$ match very well near SFRs $\sim 1 \Msun
\invyr$.  At lower and higher SFRs, the calibration is
not as effective and the simulated SFRs depend
still strongly on resolution and box size. 
\par
Generally, higher resolution
enables the formation of lower-mass galaxies with lower SFRs, and also renders stellar
feedback more effective. As a result, at low SFRs, despite
calibration, the simulations overpredict the SFR function, and this
effect is strongest at the lowest simulated redshift and for
the simulations with the lowest resolution.  On the other hand, a
larger box size reduces the bias on the estimate of the
abundance of massive galaxies that results from missing large-scale
power in the initial conditions. This impacts the comparison at high
SFRs, where the abundance of star-forming galaxies is also affected by
cosmic variance. The comparison of the
simulated and observed SFR functions 
at large SFRs may also be affected by uncertain
dust corrections and by feedback from massive black 
holes (e.g., \citealp{Feng2016}), both of which our simulations ignore.
\par
Negative feedback from photoheating reduces star formation in low-mass
galaxies, as discussed above. One may expect this to imprint
an observable change in the slope of the SFR function at low SFRs,
which correspond to low-mass halos (e.g.,
\citealp{Barkana2006}; \citealp{Munoz2011}; \citealp{Mashian2013}; \citealp{Wise2014}). However, strong
stellar feedback may mask the impact of photoheating on the SFR
function, because it also strongly reduces the normalization of the
SFR function (PSD15; see \citealp{Gnedin2014b} for an
alternative mechanism that reduces the signature of photoheating). In our
simulations, the turnover in the slope of the SFR function at low SFRs
is mostly a numerical artefact caused by the lack of resolution and low-temperature
coolants, such as molecular hydrogen.
\par
Figure~\ref{fig1a} shows that the agreement between simulated and observed SFR functions 
carries over to the comparison between the evolution of the simulated SFR 
density contributed by observationally accessible galaxies with SFRs $\gtrsim 0.2
\Msun \invyr$ and the corresponding observational constraints. The
figure also shows explicitly that our simulations predict a substantial population of faint star-forming
galaxies that are currently below the detection threshold. These will
be prime targets for upcoming observations with, e.g., the
JWST (e.g., \citealp{Johnson2009}; \citealp{PawlikBromm2011};
\citealp{Zackrisson2011}; \citealp{Wise2014}). 
\par

\subsection{Reionization history}
\label{Sec:reionization} 
The left panel of Figure~\ref{fig2} shows that all simulations yield mean neutral hydrogen
fractions in agreement with estimates from observations of the
Lyman-$\alpha$ forest at $z\approx 6$ (filled triangles). Our calibrations of the SFR
function and of the subresolution escape fraction 
implies that the gas is, on average, ionized earlier than inferred 
from observations of Lyman-$\alpha$
emitters (e.g., \citealp{Tilvi2014}, \citealp{Pentericci2014},
\citealp{Schenker2014}; see  \citealp{Robertson2015} for an overview) and quasar damping wings
(\citealp{Schroeder2013}) at $z\approx 7$. However, the
interpretation of these observations is strongly model-dependent
which makes their constraining power unclear (e.g., \citealp{Dijkstra2014}; \citealp{Gnedin2014b}). Note that
at higher resolution, reionization is more extended. This is most likely
because at higher resolution, the abundance of low-luminosity sources, as well as that of absorbers of ionizing radiation, is increased. The right panel 
of Figure~\ref{fig2a} shows that the adopted calibration of the subresolution
escape fraction implies a reionization optical depth in excellent
agreement with observational constraints (\citealp{Planck2015}).
\par
One should keep in mind that our
simulations do not resolve any ionization by an early population of
star-forming minihaloes (e.g., \citealp{Abel2002}; \citealp{Bromm2002}; \citealp{Greif2008}; \citealp{Wise2014}) and also do not account for
ionization by non-stellar sources, such as, e.g., accreting black holes (e.g.,
 \citealp{Shull2008}; \citealp{Haiman2011}; \citealp{Jeon2014}). A
 significant contribution from these sources to the total amount of
 ionizing radiation during reionization will require us to reduce the
subresolution escape fraction in order to retain the match between
simulated and observed reionization optical depths (e.g.,
\citealp{Ahn2012}; \citealp{Wise2014}). This might delay
reionization as these very low-mass sources
are unlikely to continue to contribute substantially during the later stages of
reionization, and thereby improve the match between simulated and observed neutral
hydrogen fractions at $z > 6$. For reference, a simulation
at the resolution of our reference simulation but with subresolution
escape fraction $f_{\rm esc}^{\rm subres} = 0.3$ matches most observational constraints on the
neutral fraction, while retaining
consistency with the observed electron scattering optical depth
(dotted curve in Figure~\ref{fig2}; for computational efficiency, this simulation was
carried out in a smaller box of size $12.5 \Mpch$).
\par

\subsection{Build-up of the ionizing background }
The left panel of Figure~\ref{fig2a} shows the build-up of the
background of hydrogen ionizing radiation that accompanies the
reionization of the gas.  Immediately apparent is that the simulated
volume-weighted mean photoionization rates at $z\approx 6$ are substantially larger
than the photoionization rate inferred from current observations
(e.g., \citealp{Wyithe2011}; \citealp{Calverley2011}). This could 
indicate that our adopted subresolution escape fractions are too high
and result in reionization occuring too early. Indeed, a
simulation at the resolution of our reference simulation but with
a lower subresolution escape fraction $f_{\rm esc}^{\rm subres} = 0.3$ yields
a mean photoionization rate in agreement with the observations, while
also matching the constraints on the neutral fraction and the optical
depth (dashed curves in Figure~\ref{fig2}).
\par
However, additional factors may contribute to produce a high mean
photoionization rate, and these would need to be considered upon any
attempts to recalibrate the subresolution escape fraction. For
instance, in our simulations, there is a large scatter around the mean
photoionization rate at fixed density. This scatter causes the mean photoionization
rate to be much higher than the median photoionization rate (see the long-dashed curve in the left panel of Figure~\ref{fig2a}),
and this median might be more appropriate for comparison
with the observational estimates (Rahmati et al. in
prep.). Additionally, Figure~\ref{fig2a} reveals a marked trend of
decreasing mean photoionization rates at increasing resolution near the
end of reionization, which is not
yet fully converged in our reference simulation. If this trend
continues to higher resolution, this may further help bring simulated
and observed photoionization rates in agreement.
\par
The trend of decreasing mean photoionization rates with increasing
resolution is interesting in itself. Thanks to the calibration, the
redshift of reionization in our simulations is independent of
resolution. If the redshift of reionization is mostly set by the total
amount of ionizing radiation escaping into the IGM, which our
calibration controls, the decrease with increasing resolution in
the mean photoionization rate, which scales with the total amount of
ionizing radiation escaping into the IGM and the mean free path, may
be caused primarily by the decrease in the mean free path of
ionizing photons. The mean free path is set by the abundance of
neutral hydrogen absorption systems with column densities $\sim
10^{17} \cmsi$ (e.g., \citealp{Prochaska2009}).  The right panel of
figure~\ref{fig2a} shows the column density distribution of neutral
hydrogen computed as described in \cite{Rahmati2013}. The abundance of
absorbers with column densities $N_{\rm HI} \gtrsim 10^{17} \cmsi$
grows significantly with increasing resolution between our
low-resolution and reference simulation. However, other factors may
play a role, such as, e.g., the increased duration of reionization at
higher resolution seen in the left panel of Figure~\ref{fig2}. We will
analyze this in more detail in subsequent work (Rahmati et al. in
prep.).
\par

\begin{figure}
  \begin{center}

   \includegraphics[width=0.49\textwidth,clip=true, trim=-10 20 20 30,
      keepaspectratio=true]{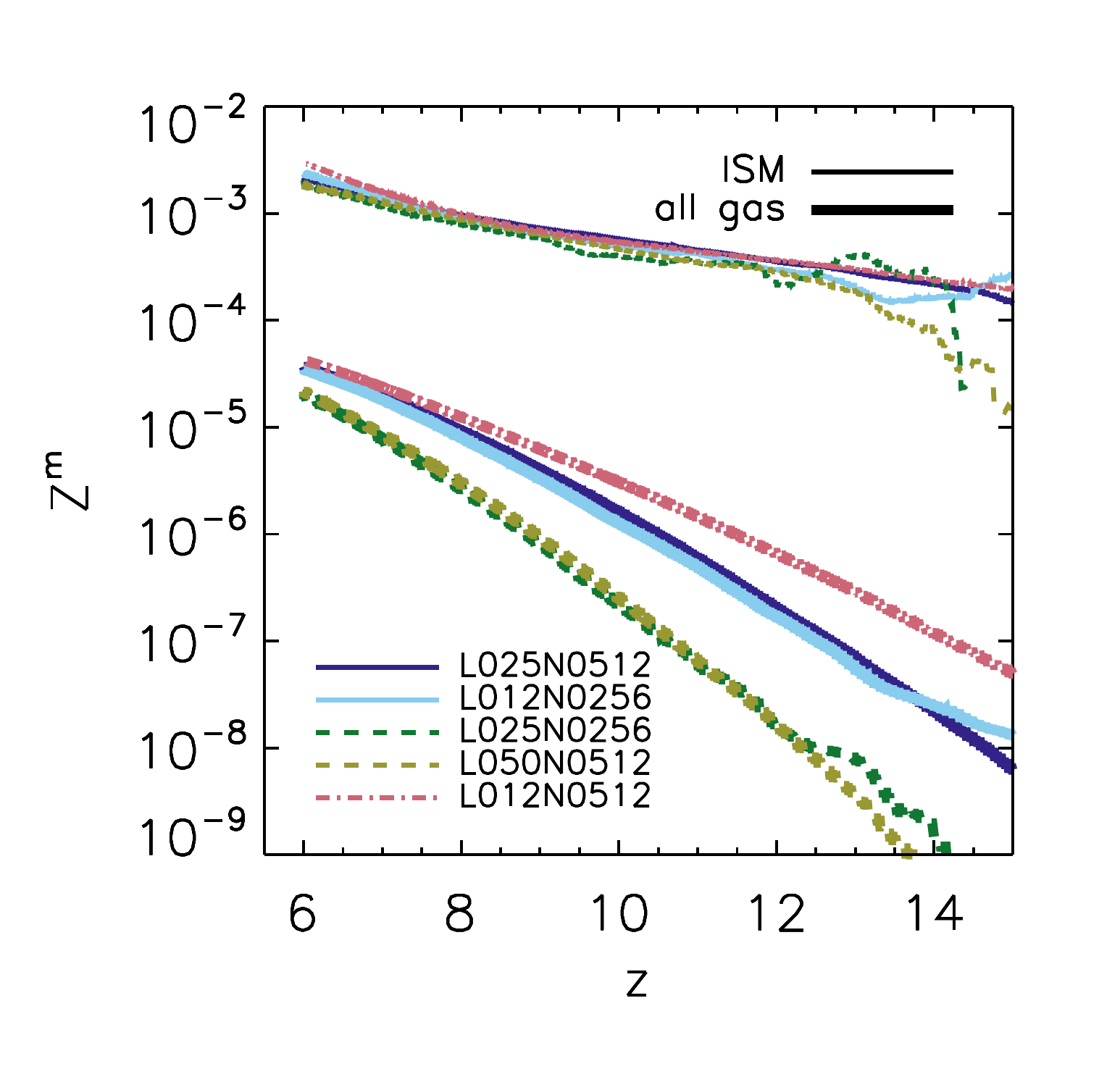}

\end{center}
\caption{Mean mass-weighted metallicity. Thin and thick curves show
  the metallicities of the ISM (i.e.,
      gas with densities $n_{\rm H} \ge 0.1 \cmci$) and of all gas
      irrespective of density.}

\label{fig2b}
\end{figure}

\subsection{Chemical enrichment history}
The galaxies that reionize the IGM also enrich the cosmic gas with
metals due to mass loss from stellar winds and explosions as
SNe. Constraints on the metal enrichment history may therefore help in
understanding how galaxies reionized the universe (e.g.,
\citealp{Oh2002}; \citealp{Keating2014}; \citealp{Finlator2015}; \citealp{Ferrara2015}). Note
that the outflow of metals into
the IGM can be facilitated by a prior episode of photoheating that
reduces the densities near the sites at which the metals are produced
(e.g., \citealp{Pawlik2009}; \citealp{Finlator2011};
\citealp{Jeon2014b}; \citealp{Walch2015}).
\par
Figure~\ref{fig2b} shows the mean
mass-weighted metallicity, defined as the ratio of metal mass and
total mass, of all gas and of gas in the ISM only (i.e., gas with $n_{\rm H} > 0.1 \cmci$). The
mean total gas metallicity is substantially smaller than
the metallicity of the ISM, consistent with the
small fraction of the volume filled with metals
(Figure~\ref{fig0}). The mean metallicity
of the star particles (not shown here) is very similar to that of the gas in the ISM.  At $z
= 6$, the mean mass-weighted gas metallicity is nearly converged with resolution 
in our reference simulation, reaching values $Z^{\rm m} \lesssim
5\times10^{-5}$, consistent with previous simulations (e.g.,
\citealp{Wiersma2009}; \citealp{Salvaterra2011}; \citealp{Pallottini2014}). 

\begin{figure*}
  \begin{center}

    \includegraphics[width=0.49\textwidth,clip=true, trim=-10 0 0 -10,
      keepaspectratio=true]{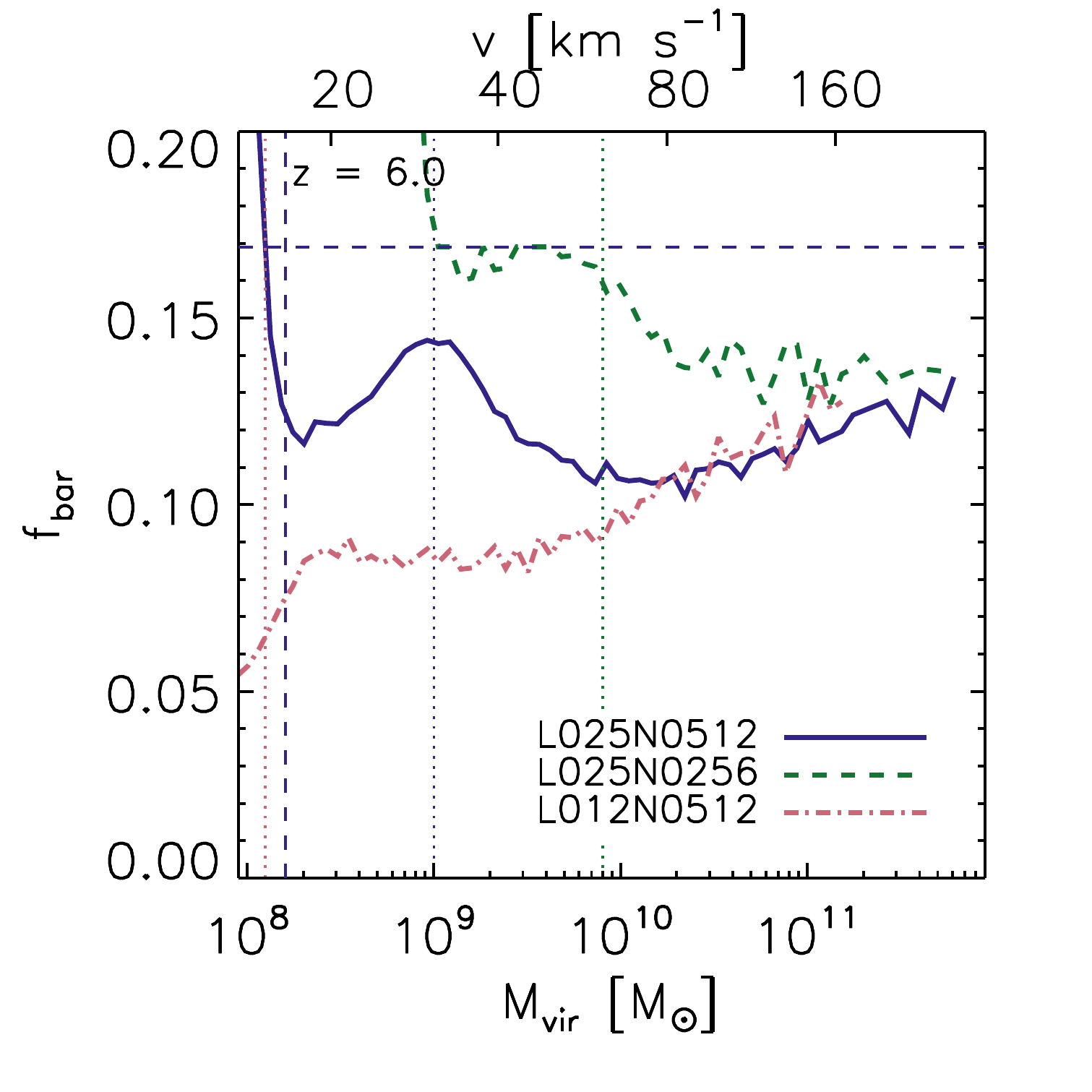}
    \includegraphics[width=0.49\textwidth,clip=true, trim=-10 0 0 -10,
      keepaspectratio=true]{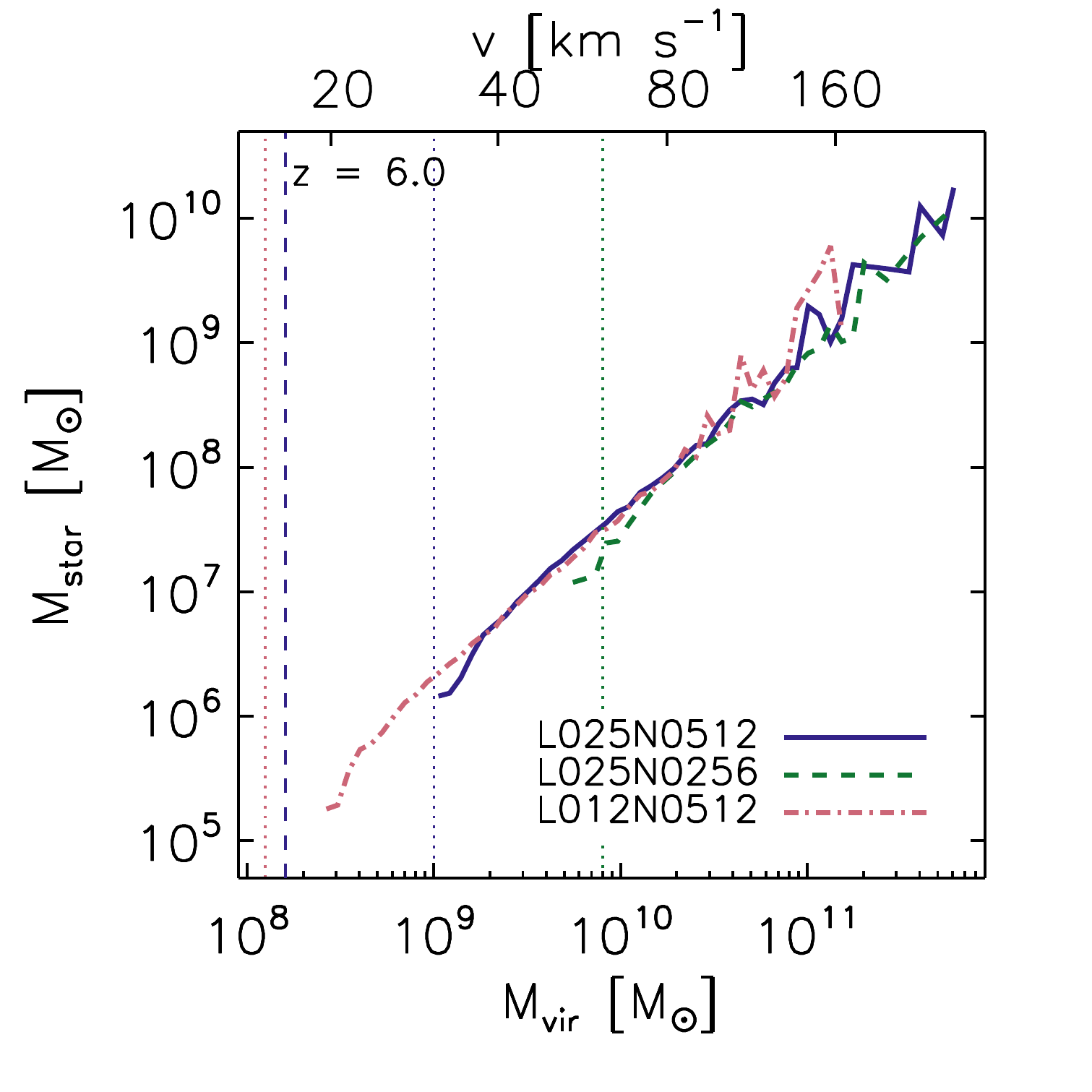}
\\
    \includegraphics[width=0.49\textwidth,clip=true, trim=-10 0 0 0,
      keepaspectratio=true]{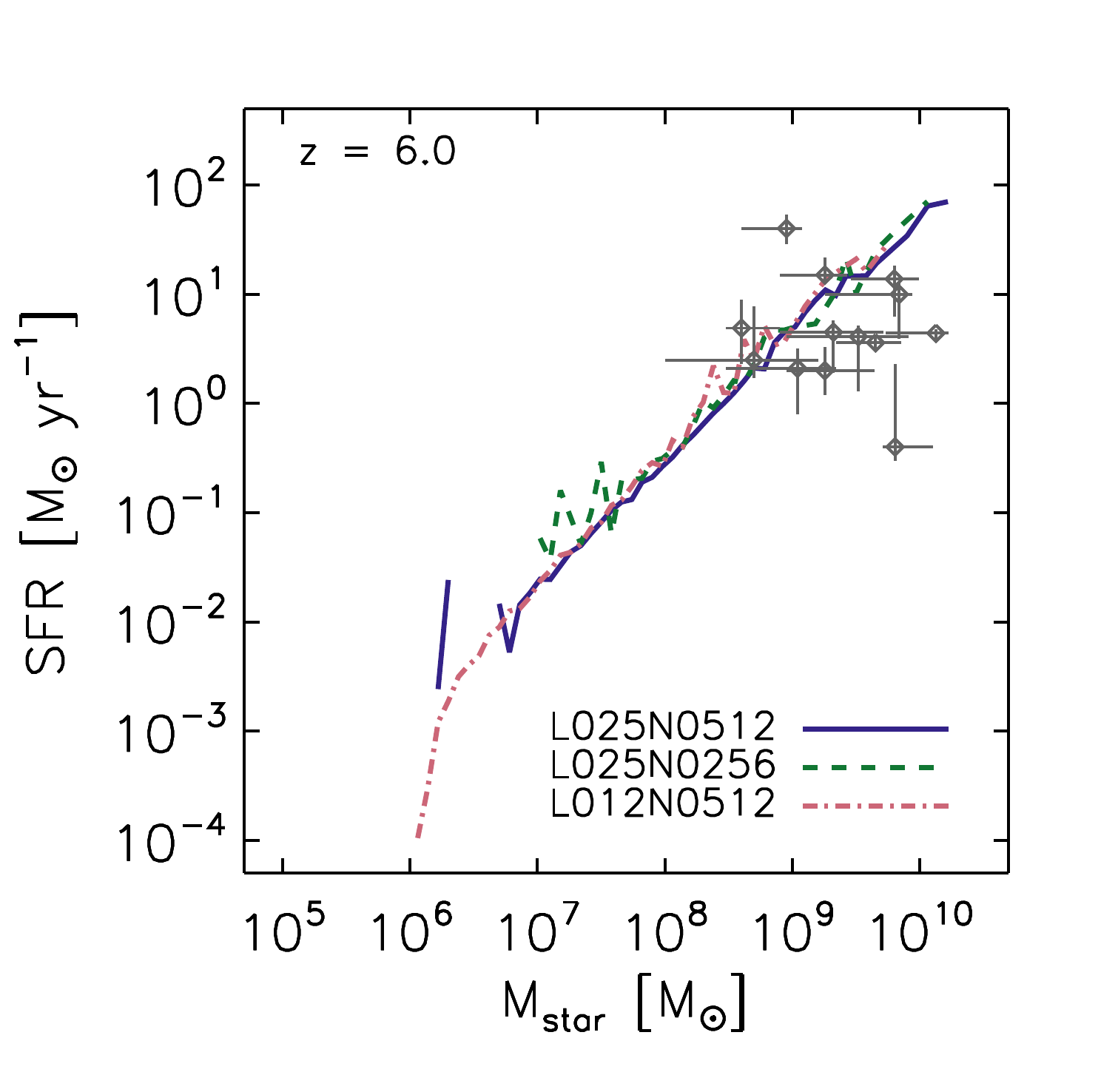}
    \includegraphics[width=0.49\textwidth,clip=true, trim=-10 0 0 0,
      keepaspectratio=true]{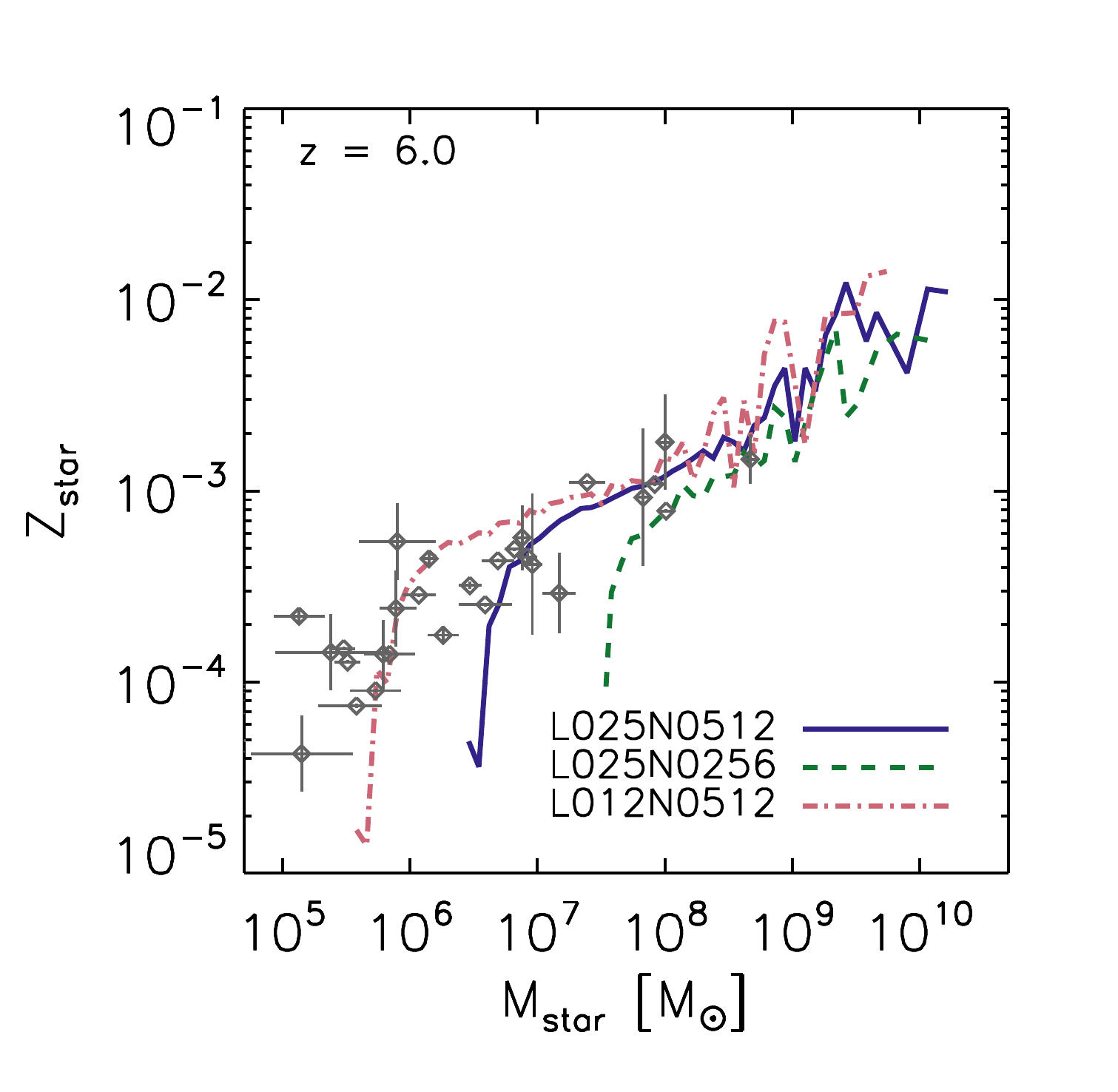}
    
  \end{center}
  \caption{Dependence on resolution of select properties of the
    simulated galaxies at z = 6. From top to bottom, left to right,
    the panels show the median baryon fraction as a function of
  virial mass, the median stellar mass as a function of virial mass, the median
  SFR as a function of stellar mass, and the mass-weighted mean stellar
  metallicity as a function of stellar mass. The vertical dotted lines
indicate the mass of 100 DM particles in the simulation of the
corresponding color, and the vertical dashed line indicates the mass
corresponding to a virial temperature of $10^4 \K$. The horizontal dashed line in the top left panel marks the cosmic baryon
fraction. The observational estimates in the bottom left panel are
from \protect\citet[their sample with redshifts $6 \lesssim z \lesssim
6.5$]{McLure2011}. The data points in the bottom right panel show the
compilation of \protect\citet{Kirby2013}  of the stellar mass-metallicity relation of Local Group dwarf galaxies.}
  \label{fig4}
\end{figure*}

\subsection{Galaxy properties}
Figure~\ref{fig4} shows select properties of the population of
galaxies at  $z = 6$ and explores their dependence on resolution in our 
calibrated set of reionization simulations. The baryon fraction (top
left panel) is strongly reduced by photoheating and stellar feedback. Stellar
feedback suppresses star formation significantly in galaxies as massive as $10^{11} \Msunh$,
consistent with previous works (e.g., \citealp{Finlator2011}; see Fig.~6 in PSD15 for an explicit demonstration). At the
reference resolution, stellar feedback is efficient only in galaxies 
more massive than $\gtrsim 10^9 \Msun$. Because photoheating only
removes substantial amounts of gas from galaxies $\lesssim 10^9 \Msun$, consistent with previous works (e.g., \citealp{Okamoto2008}; \citealp{Pawlik2009}; \citealp{Finlator2011}; \citealp{Gnedin2014b}), 
the baryon fraction near $10^9 \Msun$ is artificially increased (see Figure~6 in
PSD15 for an explicit demonstration). In the high resolution
simulation, stellar feedback is efficient down to smaller 
masses, and the baryon fraction is strongly suppressed across the
range of masses.
\par
The SFRs of the simulated galaxies (top right and bottom left panels)
are consistent with current observational estimates (\citealp{McLure2011}). The simulations
predict that the currently observable galaxies with SFRs $\gtrsim 0.2
\Msun\invyr$ are hosted by halos with virial masses $\gtrsim 10^{10}
\Msun$. Stellar feedback determines the overall
normalization of the relation between SFR and virial mass in halos with virial
masses $\gtrsim 10^9 \Msun$, while photoheating and the lack of
resolution and low temperature physics (such as molecular
cooling) are mainly responsible for the strong suppression of the SFRs
in halos with virial masses $\lesssim 10^9 \Msun$ (see Figure~6 in PSD15 for
an explicit demonstration).
\par
By $z = 6$, the most massive galaxies with
stellar masses $\gtrsim 10^9 \Msun$ have metallicities $\gtrsim
10^{-3}$ (bottom right panel). The relationship
between stellar metallicities and masses of simulated galaxies is
consistent with the stellar mass-metallicity relation of Local Group
dwarf galaxies \citep{Kirby2013}. This is interesting because those galaxies are
believed to have formed most of their stars at high
redshifts (\citealp{Weisz2014}).

\section{Summary}
We have introduced Aurora, a new suite of cosmological
radiation-hydrodynamical simulations of galaxy formation and
reionization. Aurora uses accurate 
radiative transfer (RT) at
the native, spatially adaptive resolution of the hydrodynamic
simulation. The Aurora
simulations track the stellar feedback from the explosion of stars as
supernovae (SNe) and also account for the enrichment of the universe with
metals synthesized in the stars. The reference simulation, which makes
use of a box of size $25 \cMpch$ and contains $512^3$ dark matter and
$512^3$ baryonic particles and therefore resolves atomically cooling
haloes with virial temperatures $\gtrsim 10^4 \K$ with $\gtrsim 10$ DM particles, is accompanied by
simulations in larger and smaller boxes, with, respectively, lower and higher
resolution to investigate numerical convergence.  While our most
demanding simulation, which uses $2\times 1024^3$ particles in a $100 \cMpch$ box, was stopped at $z=8.4$, all other simulations were run down to $z=6$.
\par
The Aurora simulations are designed to yield star formation rate (SFR)
functions in agreement with observations of galaxies at $z \approx 7$ and to complete
reionization at $ z \approx 8.3$, consistent with measurements of the electron scattering optical
depth, independently of the box size and
the resolution. We accomplished this by calibrating the stellar feedback
efficiency, which is the main subresolution parameter impacting the
SFR function over the range of observed SFRs, and the subresolution escape fraction,
which is the main subresolution parameter impacting the redshift of
reionization. The calibration is motivated by the fact that current
cosmological simulations lack both the physics and the resolution to
reliably simulate the structure of the ISM in galaxies. The latter
determines how much of the energy injected by massive stars drives
galactic outflows reducing the SFR, and how many of the emitted ionizing photons escape into the
IGM.
\par
While the main aim of this work was to describe the design and calibration of the simulations, we have also reported results from an initial analysis. We found that among Aurora simulations and after reionization, there is a strong decrease of the mean photoionization rate with increasing resolution. This coincides with an increase in the abundance of small-scale optically thick HI absorbers in the IGM. Given the large scatter in the photoionization rates in the Aurora simulations, the predicted IGM photoionization rates are consistent with the existing observational constraints at $z \approx 6$. The fact that the radiative transfer technique used to carry out Aurora is spatially adaptive, was critical for enabling us resolve the absorption of ionizing radiation on small scales  in radiation-hydrodynamical simulations of cosmological volumes. In future works, we will exploit this unique feature of the Aurora simulations to study different properties of small scale absorbers and radiation field, and their evolutions.

\par

\section*{Acknowledgments}
We are grateful to Volker Springel for letting us use GADGET
and the halo finder Subfind. We are grateful to Volker Bromm for letting us use
his chemistry solver. We thank Milan Raicevic for his contributions in
an early, preparatory phase of this project, and we thank Gunjan Bansal for
helpful discussions.  AHP thanks the Max Planck Institute
for Astrophysics (MPA) for its hospitality during the work on this paper. Computer resources for this project have been
provided by the Gauss Centre for Supercomputing/Leibniz Supercomputing Centre under grant:pr83le (PI Pawlik). We further
acknowledge PRACE for awarding us access to resource SuperMUC based in Germany
at LRZ Garching (proposal number 2013091919, PI Pawlik). Some of the simulations
presented here were run on Odin at the Rechenzentrum Garching (RZG) and the
Max Planck Institute for Astrophysics (MPA) and on Hydra at the
Leibniz-Rechenzentrum (LRZ) in Garching. This work was supported by a grant from the Swiss National Supercomputing Centre (CSCS) under project ID s613 (PI Rahmati). This work was sponsored with
financial support from the Netherlands Organization for Scientific Research
(NWO). We also
benefited from funding from the European Research Council under the
European Unions Seventh Framework Programme (FP7/2007-2013) / ERC Grant
agreement 278594-GasAroundGalaxies. AHP received funding from the European
Union's Seventh Framework Programme (FP7/2007-2013) under grant agreement
number 301096-proFeSsoR.

\label{lastpage}

\end{document}